\documentclass[twocolumn,aps,pr,superscriptaddress,nofootinbib,10pt]{revtex4-2}
\usepackage{amsmath,amssymb}
\usepackage[dvipdf,dvips]{graphicx}
\usepackage{color}
\usepackage{hyperref}
\usepackage{url}
\usepackage{slashed}
\usepackage{subfigure}
\usepackage{subcaption}
\usepackage[usenames,dvipsnames]{xcolor}
\usepackage{amsmath}
\usepackage{amsfonts}
\usepackage{float} 
\usepackage{amssymb}
\usepackage{epsfig}
\usepackage{graphics}
\usepackage{euscript}
\usepackage{slashed}
\usepackage{epstopdf}
\usepackage[utf8]{inputenc}
\allowdisplaybreaks
\usepackage[normalem]{ulem}
\usepackage{pifont}
\usepackage{dsfont}
\usepackage{MnSymbol}
\usepackage{verbatim}
\usepackage{graphicx}
\usepackage{latexsym}
\usepackage{courier}
\usepackage{cancel}

\hypersetup{
colorlinks=true,
citecolor=blue,
citebordercolor=red,
linktoc=all,
linkcolor=blue,
urlcolor=blue
}

\def \s{\mathsf{s}}

\def \and{\textmd{and}}

\def \be{\begin{equation}}
\def \ee{\end{equation}}

\def \bea{\begin{eqnarray}}
\def \eea{\end{eqnarray}}

\graphicspath{{./figs/}}
\newbox{\ORCIDicon}
\sbox{\ORCIDicon}{\large
                  \includegraphics[width=0.8em]{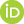}}

\begin{document}

\title{Charting the different phases of Yang-Mills-Chern-Simons-Higgs theories}

\author{Daniel O. R. Azevedo\,\href{https://orcid.org/0009-0000-0633-654X}{\usebox{\ORCIDicon}}}\email{azevedo.dor@gmail.com}
\affiliation{Instituto de F\'isica, Facultad de Ingenier\'ia, Universidad de la Rep\'ublica, J. H. y Reissig 565, 11000 Montevideo, Uruguay}

\author{Gustavo P. de Brito\,\href{https://orcid.org/0000-0003-2240-528X}{\usebox{\ORCIDicon}}} 
	\email{gp.brito@unesp.br}
	\affiliation{Universidade Estadual Paulista (UNESP), Faculdade de Engenharia e Ci\^encias de Guaratinguet\'a,
Av. Dr. Ariberto Pereira da Cunha, 333, 12516-410, Guaratinguet\'a, SP, Brazil }
    
\author{Philipe De Fabritiis\,\href{https://orcid.org/0000-0001-5455-6889}{\usebox{\ORCIDicon}}}\email{pdf321@cbpf.br}
\affiliation{CBPF - Centro Brasileiro de Pesquisas F\'isicas, \\ Rua Dr. Xavier Sigaud 150, 22290-180, Rio de Janeiro, Brazil}

\author{Antonio D. Pereira\,\href{https://orcid.org/0000-0002-6952-2961}{\usebox{\ORCIDicon}}} \email{adpjunior@id.uff.br}
\affiliation{Instituto de F\'isica, Universidade Federal Fluminense, Campus da Praia Vermelha, Av. Litor\^anea s/n, 24210-346, Niter\'oi, RJ, Brazil}

\begin{abstract}
We explore Yang-Mills-Chern-Simons theories coupled to a Higgs-like field in the fundamental representation of $SU(2)$ quantized in linear covariant gauges in three Euclidean dimensions. We analyze the modifications in the analytic structure of the gluon propagator due to the elimination of infinitesimal Gribov copies. The interplay between the Higgs, the Chern-Simons and Gribov mass parameters is investigated. Two different phases are identified: a confining one, where all poles are complex, and a deconfined one, where would-be physical gluon excitation can appear. Unlike previous works, the Gribov parameter is consistently fixed by its gap equation as a function of the other mass parameters and the gauge coupling. This imposes a constraint in the parameter space and makes transparent how the competition of the mass parameters affects the relevance of the Gribov parameter for the characterization of the spectrum of the theory.
\end{abstract}

\maketitle

\section{Introduction \label{Sec:Intro}}

Yang-Mills (YM) theories are at the heart of the Standard Model (SM) of Particle Physics and have been essential for understanding the elementary particles and their fundamental interactions.~The excellent agreement between the SM predictions and experimental data from high-energy scatterings in particle colliders over the decades makes it  a very robust pillar of our fundamental understanding of Nature.

In particular, for Quantum Chromodynamics (QCD), which describes the strong interactions between quarks and gluons, the success of perturbative calculations in high-energy processes is related to the property of asymptotic freedom~\cite{Gross:1973id,Politzer:1973fx}, which guarantees that the gauge coupling decreases for increasing energy, making perturbation theory reliable in the ultraviolet (UV) regime. On the other hand, the infrared (IR) regime is strongly coupled, making the standard perturbative techniques unreliable for sufficiently low energies. A daunting challenge of the IR regime of QCD is, for instance, to understand the mechanism that drives confinement~\cite{Greensite:2011zz, Brambilla:2014jmp}, which traps quarks and gluons inside hadrons.

Contrary to the gauge-invariant formulation of the lattice approach, in  continuum spacetime the paradigmatic picture of QCD described above is deeply rooted in the Faddeev-Popov (FP) action, obtained from the functional integral quantization of YM theories in conjunction with the FP procedure to fix the gauge~\cite{Faddeev:1967fc}. However, the existence of the so-called Gribov copies~\cite{Gribov:1977wm,Singer:1978dk} jeopardizes the FP procedure, rendering it ill-defined in the IR regime. The existence of such configurations defines what is known as the Gribov problem, see, e.g., \cite{Vandersickel:2012tz,Sobreiro:2005ec}. It is not a particular obstruction of a given gauge choice but a consequence of the non-trivial bundle structure of YM theories.

Given a gauge condition, the Gribov problem can be dealt with by different strategies. The most developed one leads to the so-called (Refined) Gribov-Zwanziger scenario in the Landau gauge, see \cite{Gribov:1977wm,Zwanziger:1989mf,Dudal:2007cw,Dudal:2008sp,Dudal:2011gd}. In practice, it consists in restricting the path integral to a region in field space, the Gribov region, which is free of (infinitesimal) Gribov copies\footnote{The Gribov region is not free of Gribov copies since it still contains those generated by finite gauge transformations~\cite{vanBaal:1991zw}. There exists a region which is completely free of Gribov copies, the Fundamental Modular Region, but a practical implementation of the restriction of the path integral to such a region is still an open problem.}.~The restriction is effectively implemented by a local and renormalizable action that takes into account further non-perturbative effects such as the formation of dimension-two condensates. In the following, we highlight the main aspects of such a procedure.

The infinitesimal Gribov copies are associated with normalizable zero-modes of the FP operator. Thus, to avoid such copies, one could simply restrict the path integral to the region where the FP operator is positive-definite, which corresponds to the so-called Gribov region. By definition, it is free of infinitesimal Gribov copies. This statement makes sense as long as the FP operator is Hermitian, which is true in the Landau gauge. To implement such restriction in practice, one can  use the intrinsic relation between the FP operator and the ghost propagator, and impose that the ghost propagator does not develop poles, guaranteeing that the FP operator will not develop zero modes. This is known as the no-pole condition proposed by Gribov~\cite{Gribov:1977wm}. Of course, it is important to be sure that such restriction does not leave out any physical information. Indeed, it was proven later that the Gribov region is bounded in all directions in field space, it is convex, and all gauge orbits cross it at least once~\cite{DellAntonio:1991mms}. The boundary of the Gribov region is known as the Gribov horizon.

The restriction to the Gribov region can also be implemented by the introduction of a non-local term known as the Horizon function~\cite{Zwanziger:1988jt} in the gauge-fixed action. It can be localized by the introduction of suitable auxiliary fields \cite{Zwanziger:1989mf}, resulting in the so-called Gribov-Zwanziger (GZ) action. This action is local, renormalizable to all orders, and effectively implements the restriction to the Gribov horizon, thereby removing all infinitesimal gauge copies. The GZ action predicts a gluon propagator that vanishes at zero momentum and an enhanced ghost propagator in the IR. Those properties are in disagreement with the behavior observed in lattice simulations~\cite{Sternbeck:2007ug,Cucchieri:2007md, Cucchieri:2007rg,Bornyakov:2008yx,Bogolubsky:2009dc}. There are further non-perturbative effects that must be taken into account in the GZ framework, such as the dynamical formation of condensates in the IR. These improvements led to the RGZ action~\cite{Dudal:2007cw,Dudal:2008sp,Dudal:2011gd}, which provides tree-level gluon and one-loop ghost propagators in good agreement with lattice data, see, e.g., \cite{Cucchieri:2011ig,Cucchieri:2016jwg}. More recently, the gluon propagator in the RGZ framework was computed at one-loop order and the qualitative features are preserved, see \cite{deBrito:2024ffa}. The behavior of one-loop corrected matter-fields propagators in the presence of the Gribov horizon and their comparison with lattice simulations were explored in \cite{deBrito:2023qfs,deBrito:2025nvl}. At this stage, a comment is in order: The comparison between the RGZ computations and lattice simulations is achieved by fitting the mass-parameters that appear in the theory. In principle, those parameters can be fixed dynamically, but this requires solving sufficiently complicated gap equations that might not provide good qualitative results at leading-order in perturbation theory. One of the most important and interesting open problem in this scenario is precisely the determination of those massive parameters self-consistently. First steps were given in \cite{Dudal:2005na,Dudal:2011gd,Dudal:2019ing}.

The RGZ gluon propagator admits complex poles, which can be interpreted as an evidence for the unphysical nature of the corresponding excitation, a property that is compatible with confinement. However, the inclusion of a new mass-parameter through the coupling of a scalar field can change such a picture. This is due to the competition between the corresponding mass-parameters in the propagator, see, e.g.,~\cite{Capri:2012jhc,Capri:2014jhb, Capri:2012cr,Capri:2013gha}. Interestingly enough, it has been shown that the YM-Higgs model, with the Higgs in the fundamental representation, can exhibit a smooth transition between the Higgs and confining phases\footnote{It should be emphasized that the term "phase" is used here in an operational rather than a strict thermodynamic sense. As the parameters of the theory are varied, the analytic structure of the propagator changes, giving rise to regions of parameter space where the poles are either real or complex. Motivated by the usual interpretation of these two situations in terms of physical and confined excitations, respectively, we refer to these regions as "phases". This terminology is common in the related literature; see, e.g., ~\cite{Capri:2012jhc,Capri:2014jhb, Capri:2012cr,Capri:2013gha}.} for a certain region of the parameter space of the Higgs vacuum expectation value ($v$) and the gauge coupling ($g$), called the analyticity region~\cite{Fradkin:1979fef}. 

In three spacetime dimensions, it is also possible to generate mass for the gauge fields in harmony with gauge invariance through a topological Chern-Simons (CS) term, which also alters the pole structure of the gluon propagator~\cite{Canfora:2014qwe,Gomez:2016egs, Ferreira:2020hcj, Felix:2021eoq,Ferreira:2021ksd}, allowing for deconfined excitations. The introduction of a CS mass term is appealing since it is a local term compatible with infinitesimal gauge invariance~\cite{Deser:1982fsr,Deser:1982kay}, which does not require the introduction of new fields in the action. Under finite gauge transformations, a restriction over the topological mass value must be imposed, which we do not require here. The CS term, however, is not parity-invariant, which can be seen by the presence of the totally anti-symmetric Levi-Civita symbol $\epsilon_{\mu\rho\nu}$.~Beyond that, both the pure CS and the YMCS theories are finite~\cite{Delduc:1990je,Giavarini:1992xz,DelCima:1997pb,DelCima:1998ur,Azevedo:2024cov}. Recently, this property was verified for YMCS taking into account the presence of Gribov copies~\cite{Azevedo:2025yal} in linear covariant gauges.

The goal of this work is to investigate the analytic structure of the gluon propagator and the different possible phases in YMCS theories coupled to a Higgs field in the fundamental representation of the $SU(2)$ gauge group when taking into account the effects of (infinitesimal) Gribov copies. This is carried out in linear covariant gauges. In particular, we investigate the analytical structure of the gluon propagator, which provides hints about the confining and deconfining phases of the theory. The key motivation for the present work is that in three-dimensions, due to the finiteness of the theory, it is possible to envision a solution of (some) gap equations and, instead of treating all mass parameters independently, they can be dynamically intertwined. In such better controlled environment we can learn potential key features that will also appear in four dimensions.

The paper is organized as follows. In Sec.~\ref{TheorFrame} we discuss the setup: We introduce the YMCS model coupled to a Higgs field and compute the tree-level gluon propagator in Subsec.~\ref{Sec:YMCSH}; we discuss the elimination of Gribov copies through the introduction of the Horizon function in Subsec.~\ref{Sec:InfCopies}; the local and BRST-invariant formulation of the GZ action is presented in Subsec.~\ref{sec:brst}. Sec.~\ref{sec:gapeq} is dedicated to the gap equation analysis. The analytical properties of the gluon propagator are investigated in Sec.~\ref{sec:irprop}, where we also discuss the confining/deconfining phases of the theory. Finally, we collect our concluding remarks in Sec.~\ref{sec:conc}.

\section{Theoretical framework} \label{TheorFrame}

\subsection{Yang-Mills-Chern-Simons-Higgs action\label{Sec:YMCSH}}

Let us consider a $SU(2)$ Yang-Mills theory with a Chern-Simons term, minimally coupled to a scalar field in the fundamental representation, described by the following action in three-dimensional Euclidean spacetime:  
\begin{align}
	S_{1} = S_{\rm YM} + S_{\rm CS} + S_{\rm \phi}\,,
\end{align}
with 
\begin{align}
	S_{\rm YM} &= \frac{1}{4}\int {\rm d}^3x~ F_{\mu\nu}^a F_{\mu\nu}^a\,,  \\
     S_{\rm CS} \! &= \! -iM \!\!\! \int \!\!\! {\rm d}^3x \,  \epsilon_{\mu\rho\nu}\left(\frac{1}{2}A_\mu^a \partial_\rho A_\nu^a \!+\! \frac{g}{3!}f^{abc} A_\mu^a A_\rho^b A_\nu^c\right)\,, \\
     S_{\rm \phi} &= \int {\rm d}^3x \left[\vert D_\mu \phi \vert^2 + \Lambda \left(\phi^\dagger\phi -\frac{v^2}{2}\right)^2 \right]\,.
\end{align}
In the pure YM part, $F_{\mu\nu}^a = \partial_\mu A^a_\nu - \partial_\nu A^a_\mu +gf^{abc}A^b_\mu A^c_\nu$ is the field strength, $g$ is the gauge coupling and $f^{abc}$ are the structure constants of $SU(2)$. The $\epsilon_{\mu\rho\nu}$ is the totally anti-symmetric Levi-Civita symbol and $M$ is the CS mass parameter. This topological term provides a mass to the gauge boson while retaining gauge invariance under infinitesimal gauge transformations~\cite{Deser:1982fsr,Deser:1982kay}\footnote{It is well-known that the CS term can also be made invariant under finite gauge transformations by imposing constraints on the mass parameter $M$. However, since we will only consider Gribov copies associated with infinitesimal gauge transformations, this kind of constraint will not be considered here.}. The covariant derivative in the fundamental representation is given by $\left(D_\mu \phi\right)^i = \partial_\mu \phi^i -ig(T^a)^{ij}A^a_\mu \phi^j$, where $(T^a)^{ij}$ are the generators of the $SU(2)$ gauge group in the fundamental representation. We employ the shorthand notation $\vert D_\mu \phi \vert^2 = (D^{ij}_\mu\phi^j)^\dagger (D^{ik}_\mu \phi^k)$. The Higgs self-coupling is denoted as $\Lambda$. The scalar potential $V(\phi) = \Lambda \left(\phi^\dagger\phi -\frac{v^2}{2}\right)^2$, induces a non-trivial vacuum expectation value (VEV) for the scalar field, given by $\langle\phi\rangle = \frac{1}{\sqrt{2}}\left(\begin{matrix}
		0\\
		v
	\end{matrix}\right)$. This will drive the theory into the Higgs phase, where the gauge boson acquires a mass through the Higgs mechanism, given by $m_\phi^2 = \frac{g^2v^2}{4}$. It is clear that the gauge boson will have two different contributions for its mass: one coming from the CS term and the other from the Higgs sector. The interplay between the different mass parameters present in this model will lead to interesting features in the gauge-field propagator, see \cite{Gomez:2016egs}.  

The functional integral quantization can be performed through the well-known FP procedure. Considering linear covariant gauges, the FP action can be written as
\begin{align}
	 	 S_{2} = \int {\rm d}^3x \left(b^a\partial_\mu A_\mu^a - \frac{\alpha}{2}b^a b^a + \bar{c}^a \partial_\mu D_\mu^{ab}c^b \right),
\end{align}
where $\left(c^a, \bar{c}^a\right)$ are the FP ghosts, $D_\mu^{ab} = \delta^{ab} \partial_\mu - g f^{abc}A_\mu^c$ is the covariant derivative in the adjoint representation, $\alpha$ is a non-negative gauge parameter, and $b^a$ is the Nakanishi-Lautrup field. Therefore, the gauge-fixed action for the Yang-Mills-Chern-Simons-Higgs (YMCSH) model is
\begin{align}\label{inv_action}
	S_{\rm YMCSH} = S_1 + S_{2}.
\end{align}
The action obtained after the FP procedure is not gauge-invariant, but enjoys a symmetry under BRST transformations, defined by
\begin{equation}
\begin{aligned}
	&\s A^a_\mu = -D_\mu^{ab}c^b\,, & &\s c^a = \frac{g}{2}f^{abc}c^b c^c\,,\\
	&\s\bar{c}^a = b^a\,, & &\s b^a = 0\,,\\
	&\s\phi^i = igc^a(T^a)^{ij}\phi^j\,,
\end{aligned}
\end{equation}
with $\s^2 = 0$. The tree-level gauge-field propagator is
\begin{align}\label{gluonpropUV}
		\langle A^a_\mu(k) A^b_\nu(-k) \rangle_0 &=  \delta^{ab}\frac{k^2 + m_\phi^2}{(k^2 + m_\phi^2)^2 + M^2k^2}\left(\mathcal{P}^T_{\mu\nu}(k)\right. \nonumber \\
		&+ \left.\frac{M}{k^2 + m_\phi^2}\epsilon_{\mu\rho\nu}k_\rho\right) + \delta^{ab}\frac{\alpha}{k^2}\frac{k_\mu k_\nu}{k^2}\,,
\end{align}
where $\mathcal{P}^T_{\mu\nu}(k) \equiv \delta_{\mu\nu} - \frac{k_\mu k_\nu}{k^2}$ is the transverse projector. The CS term introduces a transverse and parity-odd part to the propagator proportional to the Levi-Civita symbol.

\subsection{Eliminating Gribov copies} \label{Sec:InfCopies}

The action~\eqref{inv_action} corresponds to the YMCSH theory gauge fixed in linear covariant gauges by means of the FP procedure. However, as first pointed out by Gribov in 1978~\cite{Gribov:1977wm}, the gauge-fixing procedure described above does not fix the gauge completely: the gauge condition does not pick only one representative per gauge orbit. Gauge-field configurations which are related by a gauge transformation and satisfy the gauge condition, the so-called Gribov copies, exist. Although this problem was first investigated in the Landau gauge, it was shown that it afflicts a large class of gauge-fixing conditions~\cite{Singer:1978dk}. The Gribov copies associated with infinitesimal gauge transformations are related to the zero modes of the FP operator, jeopardizing the FP procedure and asking for its improvement. 

Gribov himself proposed a solution to this problem: the functional integral should be restricted to a region in which the FP operator is positive-definite, the Gribov region. It was rigorously proven later that the Gribov region has at least one representative per gauge orbit~\cite{DellAntonio:1991mms}, guaranteeing that such a restriction only eliminates redundant configurations. This restriction was implemented by Gribov at leading order through the so-called no-pole condition, and at all orders by Zwanziger~\cite{Zwanziger:1989mf} following another route, by means of the horizon function. Interestingly enough, by extending the no-pole condition to all orders in perturbation theory, one is led to the same result found by Zwanziger with the horizon function~\cite{Capri:2012wx}, a substantial non-trivial check of its validity. Such developments were achieved in the Landau gauge.

In a nutshell, the restriction to the Gribov region can be implemented (in the Landau gauge) by adding a horizon term to the gauge-fixed action, such that\footnote{We write the results directly in three dimensions since the restriction to the Gribov region can be worked out for generic $d$.}
\begin{align}
    S_{\rm GZ}^{\rm nl} = S_{\rm YMCSH}^{(\alpha \to 0)} + \gamma^4 \left[ H(A) - 3 V (N^2-1)\right]\,, 
\end{align}
where $V$ is the spacetime volume, $N=2$ since we are working with gauge group $SU(2)$, and the horizon function is defined by 
\begin{align}\label{horfunc}
	H(A) \!=\! g^2 \!\! \int \!\! {\rm d}^3x {\rm d}^3y \, f^{abc}A_\mu^b(x) \left[\mathcal{M}^{-1}\right]^{ad}_{xy} f^{dec}A_\mu^e(y)\,.
\end{align}
The action $S_{\rm GZ}^{\rm nl}$ is the so-called Gribov-Zwanziger (GZ) action. The Gribov parameter $\gamma$ is not free: it is fixed  in a self-consistent way by means of the horizon condition/gap equation
\begin{align}
    \langle H(A) \rangle = 3V(N^2-1)\,,
\end{align}
where the expectation value is taken with respect to the GZ action.

The horizon function written above is non-local since it depends on the inverse of the FP operator. However, it is possible to rewrite it in a local fashion by introducing two pairs of auxiliary fields: a commuting pair $(\varphi,\bar{\varphi})_\mu^{ab}$, and an anti-commuting one $(\omega,\bar{\omega})_\mu^{ab}$. Therefore, the GZ action written in its local form in the Landau gauge is
\begin{align}
    S_{\rm GZ} &= S_{\rm YMCSH}^{(\alpha \to 0)} \! + \!\! \int \!\! {\rm d}^3x \left[-\bar{\varphi}_\mu^{ac} \mathcal{M}^{ab} \varphi_\mu^{bc} + \bar{\omega}_\mu^{ac} \mathcal{M}^{ab} \omega^{bc}\right] \nonumber \\
    &+  \!\!\!\int \!\!\! {\rm d}^3x \,\, g \gamma^2 f^{abc}A_\mu^a (\varphi_\mu^{bc}+\bar{\varphi}_\mu^{bc}) - \gamma^4 3 V (N^2-1)\,.
\end{align}
In the local formulation, the gap equation reads
\begin{equation}
    \frac{\partial \mathcal{E}_v}{\partial \gamma^2} = 0\,,
\end{equation}
where $\mathcal{E}_v$ is the vacuum energy.

The Gribov-Zwanziger action written above is an effective way of implementing the restriction to the Gribov region in a local and renormalizable framework. We define the BRST operator acting on the auxiliary fields as BRST doublets\footnote{The BRST transformations are: $\s\bar{\omega}_\mu^{ab} = \bar{\varphi}_\mu^{ab}$; $ \s\bar{\varphi}_\mu^{ab} = 0$ and $\s\varphi_\mu^{ab} ={\omega}_\mu^{ab}; \s{\omega}_\mu^{ab} = 0$.}. Yet the GZ action breaks BRST symmetry explicitly, although in a soft manner, since this breaking is proportional to $\gamma^2$, which vanishes in the ultraviolet regime. Nevertheless, a BRST-invariant formulation of the GZ action exists and is required to consistently extend the elimination of Gribov copies to other gauges such as linear covariant gauges. This will be discussed in the next subsection.

\subsection{A BRST-invariant implementation} \label{sec:brst}

A BRST-invariant version of the GZ action was developed in \cite{Capri:2015ixa}. The horizon function is suitably modified by the introduction of a dressed, gauge-invariant field $A^{h,a}_\mu$. In this formulation, the auxiliary localizing fields do not transform as doublets and the BRST operator is nilpotent. The resulting action implements the restriction of the path integral to a region free of infinitesimal Gribov copies in harmony with BRST invariance, see \cite{Capri:2015nzw,Capri:2016aqq,Capri:2016gut,Capri:2017bfd,Capri:2018ijg}. 

One can use the BRST-invariant formulation to extend the results obtained in the Landau gauge to  linear covariant gauges. This has been worked out for YMCS theories in \cite{Ferreira:2020hcj}. In the present case, the BRST-invariant action which effectively eliminates infinitesimal Gribov copies in YMCSH theories quantized in linear covariant gauges is given by 
\begin{align}
    S_{\rm GZ}^{\rm BRST} &= S_{\rm YMCSH} \nonumber \\
    &+  \int \! {\rm d}^3x \left[-\bar{\varphi}_\mu^{ac} \mathcal{M}^{ab}(A^h) \varphi_\mu^{bc} + \bar{\omega}_\mu^{ac} \mathcal{M}^{ab}(A^h) \omega^{bc}\right] \nonumber \\
    &+  \int \! {\rm d}^3x \,\, g \gamma^2 f^{abc}A_\mu^{h,a} (\varphi_\mu^{bc}+\bar{\varphi}_\mu^{bc}) - \gamma^4 3 V (N^2-1) \nonumber \\
    &+ \int \! {\rm d}^3x \left[\tau^a\partial_\mu A_\mu^{h,a}-\bar{\eta}^a\mathcal{M}^{ab}(A^h)\eta^b \right]\,.
\end{align}

The gauge-invariant field $A_\mu^{h,a}$ is defined as
\begin{equation}
	A_\mu^{h,a}T^a = h^\dagger A_\mu h +\frac{i}{g}h^\dagger\partial_\mu h\,,
\end{equation}
where the field $h$ can be defined in terms of a Stückelberg-like field $\xi= \xi^aT^a$ as
\begin{equation}
	h = e^{ig\xi^aT^a}\equiv e^{ig\xi}\,.
\end{equation}
The operator $\mathcal{M}^{ab}(A^h)$ is the FP operator written in terms of the dressed field $A_\mu^{h,a}$, that is,
\begin{equation}
	\mathcal{M}^{ab}(A^h) = - \partial_\mu D_\mu^{ab}(A^h) =- \partial_\mu ( \delta^{ab}\partial_\mu - gf^{abc}A_\mu^{h,c})\,.
\end{equation}
The field $\tau^a$ acts as a Lagrange multiplier, enforcing a transversality condition on the dressed field\footnote{For more details on the construction of the dressed field we refer to, e.g., \cite{Capri:2015ixa,Capri:2017abz}} $A_\mu^{h,a}$. Such a constraint requires the introduction of a Jacobian, associated with the ghosts $(\bar{\eta},\eta)^a$. The complete set of BRST-transformations which leaves the resulting action invariant is
\begin{equation}\label{brstmod}
\begin{aligned}
		&\s A^a_\mu = -D_\mu^{ab}c^b, & &\s c^a = \frac{g}{2}f^{abc}c^b c^c,\\
		&\s\bar{c}^a = b^a, & &\s b^a = 0,\\
		&\s\phi^i = igc^a(T^a)^{ij}\phi^j, & &\s h^{ij} = -igc^a(T^a)^{ik}h^{kj}\\
		&\s\bar{\omega}_\mu^{ab} = 0, & &\s\bar{\varphi}_\mu^{ab} = 0,\\
		&\s\varphi_\mu^{ab} =0, & &\s{\omega}_\mu^{ab} = 0,\\
		&\s A_\mu^{h,a}=0, & &\s\tau^a =0,\\
		&\s\bar{\eta}^a=0, & &\s\eta^a=0,\\
		&\s\xi^a = g^{ab}(\xi)c^b,
\end{aligned}
\end{equation}
with
\begin{equation}\label{sxi}
	g^{ab}(\xi) = -\delta^{ab} + \frac{g}{2}f^{abc}\xi^c -\frac{g^2}{12}f^{amr}f^{mbq}\xi^q\xi^r + O(\xi^3).
\end{equation}

The horizon function $H(A^h)$ can be recovered by integrating out all the auxiliary fields, resulting in
\begin{equation}\label{dresshorfunc}
	H(A^h) \!=\! g^2 \!\!\! \int \!\!\! {\rm d}^3x {\rm d}^3y f^{abc} \! A^{h,b}_\mu(x) [\mathcal{M}(A^h)^{-1}\!]^{ad}_{xy} f^{dec} \! A^{h,e}_\mu(y).
\end{equation}
This can be interpreted as a dressing of Eq.~\eqref{horfunc} and it has the role of restricting the functional integration over the gauge field $A^a_\mu$ to the region $\Omega_h$, defined as
\begin{equation}
	\Omega_h = \{A_\mu^a; \partial_\mu A^a_\mu = \alpha b^a\,|\,- \partial_\mu D_\mu^{ab}(A^h)>0\}.
\end{equation}

As previously discussed the Gribov parameter $\gamma$ is not free, but fixed by a gap equation, which can be obtained through a minimization of the vacuum energy in the YMCSH-GZ model, yielding ($N=2$):
\begin{equation}\label{gapeq}
	\frac{4}{3} g^2\int \frac{{\rm d}^3k}{(2\pi)^3}\frac{k^4 + m_\phi^2k^2 + \lambda^4}{(k^4 + m_\phi^2k^2 + \lambda^4 )^2 + M^2k^6} = 1,
\end{equation}
where $\lambda^4 \equiv 4g^2\gamma^4$. This gap equation fixes $\gamma$ as a function of the other parameters. Notice that there is no dependence on the gauge parameter $\alpha$ in the gap equation. This is an indication of the physical nature of $\gamma$, which can be confirmed by noting that it does not enter the GZ action as a BRST-exact term, allowing for its insertion in correlation functions of gauge-invariant operators.

The tree-level gauge propagator in the linear covariant gauge, after the restriction to the Gribov region, reads
\begin{equation}\label{propgribov}
	\begin{split}
		\langle A^a_\mu(k) A^b_\nu(-k) \rangle &=  \delta^{ab}\frac{(k^4 + m_\phi^2k^2 + \lambda^4)k^2}{(k^4 + m_\phi^2k^2 + \lambda^4)^2 + M^2k^6}\left(\mathcal{P}^T_{\mu\nu}(k)\right.\\
		&+ \left.\frac{Mk^2}{k^4 + m_\phi^2k^2 +\lambda^4}\epsilon_{\mu\rho\nu}k_\rho\right) + \delta^{ab}\frac{\alpha}{k^2}\frac{k_\mu k_\nu}{k^2}.
	\end{split}
\end{equation}
From this expression, one can see that only the transverse part receives contributions from the restriction to the Gribov region $\Omega_h$. Moreover, using the BRST-invariant formulation one can show that the longitudinal parity-preserving part of the gauge propagator is exact, not receiving contributions from higher-loop orders~\cite{Ferreira:2021ksd}.  

It is clear that the pole structure of the propagator depends on $\lambda$. The presence of such mass-parameter affects the spectrum of the theory. Furthermore, the non-trivial pole structure of the propagator is restricted to the transverse part, being the longitudinal one exact, as argued above. Therefore, we can  assume the Landau gauge ($\alpha \to 0$) to investigate the pole structure of the gauge propagator in the following, without loss of generality since this will give the very same result as one would obtain for the linear covariant gauge thanks to the gauge-parameter independence of the pole mass~\cite{Capri:2016gut}. 

In the following, we investigate the gap equation at one-loop order as defined by Eq.~\eqref{gapeq}. It is clear from Eq.~\eqref{gapeq} that the gap equation is defined by a finite expression in three dimensions. This fact greatly simplifies the analysis we present in the next section in contrast to the situation in four dimensions (for pure YM or YM-Higgs theories) where an appropriate renormalization procedure must be employed, see, e.g.,~\cite{Dudal:2005na,Dudal:2011gd,Dudal:2019ing}. This simplification makes the three-dimensional case a useful environment to understand qualitative features of the gap equation and its underlying solution which might provide useful insights to the more complicated four-dimensional case.


\section{Gap equation analysis}\label{sec:gapeq}

In order to discuss the infrared structure of the gluon propagator self-consistently, we investigate the gap equation~\eqref{gapeq} and its solution considering different scenarios in the parameter space. This allows us to obtain the parameter $\lambda$  in terms of the other parameters in a self-consistent way, respecting the constraint imposed by the theory and reducing the parameter space. This is a distinguishing feature of this work, since the Gribov parameter is mostly treated as a free parameter in the literature.

There are two mass parameters $(M, m_\phi)$ entering the gap equation~\eqref{gapeq} besides $\lambda$, which will be determined as a function of them. We will first focus on the simpler scenario where one (or both) of the massive parameters is set to zero.

\begin{figure*}[t!]
\begin{minipage}{0.45\textwidth}
  \includegraphics[width=\linewidth]{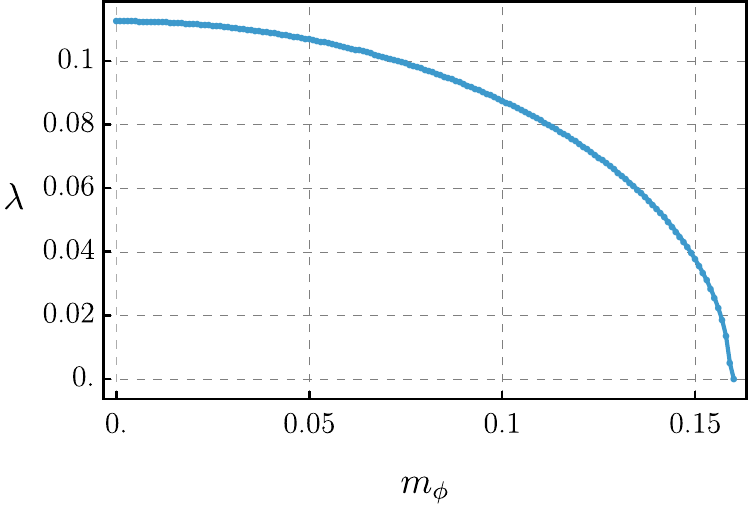}
  \caption{Behavior of the Gribov parameter $\lambda$ as a function of $m_\phi$ for $M=0$ and $g^2=1.5$.}
  \label{M0}
\end{minipage}%
\hfill 
\begin{minipage}{0.45\textwidth}
  \includegraphics[width=\linewidth]{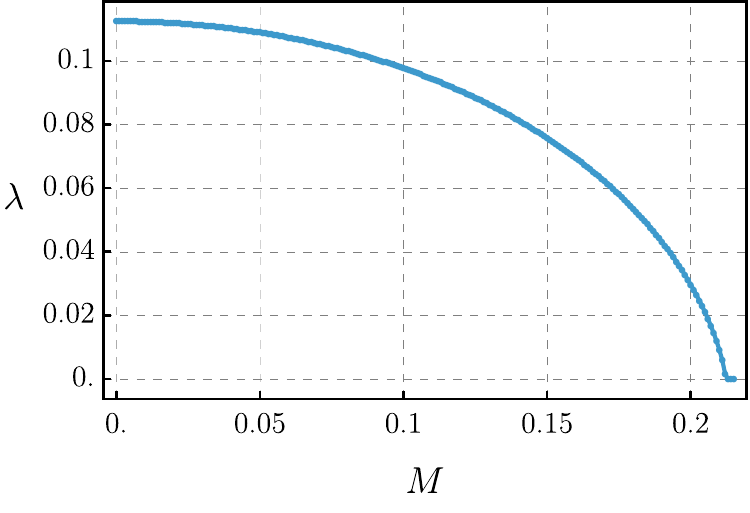}
  \caption{Behavior of the Gribov parameter $\lambda$ as a function of $M$ for $m_\phi=0$ and $g^2=1.5$.}
  \label{m_phi0}
\end{minipage}%
\end{figure*}

\subsection{Case 1: $m_\phi = 0$ and $M = 0$}

The simpler situation arises when we consider $M = m_\phi = 0$. In this case, the gap equation~\eqref{gapeq} is given by:
\begin{align}
    \frac{4}{3} g^2 \int \! \frac{{\rm d}^3k}{(2 \pi)^3} \frac{1}{k^4 + \lambda^4} = 1.
\end{align}
The above integral can be easily computed, leading to
\begin{align}
    \int \! \frac{{\rm d}^3k}{(2 \pi)^3} \frac{1}{k^4 + \lambda^4} = \frac{1}{4 \sqrt{2} \pi \lambda}
\end{align}
Thus, the gap equation can be immediately solved, yielding
\begin{align} \label{GapEqCase1}
    \lambda = \frac{g^2}{3 \sqrt{2} \pi}.
\end{align}
Therefore, considering that the dimensionless counterpart of the coupling $g$ goes to zero in the deep UV, we can immediately see that the (dimensionless) Gribov parameter goes to zero in the UV regime, confirming the expectation that the Gribov copies are not relevant in the high-energy regime, where perturbation theory holds.

\subsection{Case 2: $m_\phi \neq 0$ and $M = 0$}

Let us consider the case where $M = 0$ but $m_\phi \neq 0$. The relevant integral to be computed is

\begin{align}
    &\int \! \frac{{\rm d}^3k}{(2 \pi)^3} \frac{1}{k^4 + m_\phi^2 k^2 +\lambda^4} =  \frac{1}{2 \sqrt{2} \pi} \nonumber \\
    &\times \frac{1}{\left( \sqrt{m_\phi^2 - \sqrt{m_\phi^4 - 4 \lambda^4}} + \sqrt{m_\phi^2 + \sqrt{m_\phi^4 - 4 \lambda^4}} \right)}.
\end{align}
Thus, the gap equation reads
\begin{align}
    \left( \sqrt{m_\phi^2 - \sqrt{m_\phi^4 - 4 \lambda^4}} + \sqrt{m_\phi^2 + \sqrt{m_\phi^4 - 4 \lambda^4}} \right) = \frac{2 g^2}{3 \sqrt{2} \pi}.
\end{align}
We aim at an expression for $\lambda$ in terms of  $g$ and $m_\phi$. Thus,
\begin{align}\label{gapeqYMH}
    \lambda^2 = \frac{g^4}{18 \pi^2} - \frac{m_\phi^2}{2}.
\end{align}
Imposing $\lambda^2 > 0$ yields
\begin{align}\label{condM0}
    g^4 > 9 \pi^2 m_\phi^2.
\end{align}
Therefore, for a given $m_\phi$, we find a condition on $g$ such that there is a real solution to the gap equation. This case was first studied in~\cite{Capri:2012cr} for the Higgs field in both the adjoint and fundamental representations of the gauge group, where they arrive at the same result reported here when considering the fundamental representation case. We plot $\lambda$ as a function of $m_\phi$ in figure \ref{M0}, using $g^2 = 1.5$ as a benchmark value. 

\subsection{Case 3: $m_\phi = 0$ and $M \neq 0$}
On the other hand, if we consider $m_\phi = 0$ but $M \neq 0$, the gap equation~\eqref{gapeq} can be written as
\begin{align}\label{gapeqM}
    \int \! \frac{{\rm d}^3k}{(2 \pi)^3} \frac{k^4 + \lambda^4}{\left(k^4 + \lambda^4\right)^2 + M^2 k^6} = \frac{3}{4 g^2} 
\end{align}
We can adopt spherical coordinates and expand the above integrand using its Taylor series\footnote{This is done under the assumption that we can commute the sum and the integral, which is reasonable since the integral is finite and the corresponding series is convergent.}, obtaining
\begin{align}
   \sum_{n=0}^{\infty} \frac{(-1)^n}{2 \pi^2} \int_0^{\infty} \! {\rm d}k \, \frac{k^{6n + 2} M^{2 n}}{\left(k^4 + \lambda^4\right)^{2n+1}} = \frac{3}{4 g^2}
\end{align}
Now, we can explicitly integrate each term of the series, leading to
\begin{align}
\sum_{n=0}^{\infty} (-1)^n \frac{M^{2n} \lambda^{(-1 -2n)}}{8 \pi^2} \frac{\Gamma\left(\frac{2n+1}{4}\right) \Gamma\left(\frac{6n+3}{4}\right)}{\Gamma\left(2n+1\right)}      = \frac{3}{4 g^2} 
\end{align}
The series above is convergent and can be written in terms of a hypergeometric function as
\begin{align}\label{gapeqm0}
    \frac{1}{128 \sqrt{2} \pi \lambda^3} \left(32 \lambda^2 X_1 - 5 M^2 X_2\right) =  \frac{3}{4 g^2},
\end{align}
where \begin{align}
    X_1 &= {}_3F_2\left(\frac{1}{4},\frac{7}{12},\frac{11}{12};\frac{1}{2},\frac{3}{4};\frac{27M^4}{256\lambda^4}\right)    \nonumber \\
    X_2 &= {}_3F_2\left(\frac{3}{4},\frac{13}{12},\frac{17}{12};\frac{5}{4},\frac{3}{2};\frac{27M^4}{256\lambda^4}\right)
\end{align}
with ${}_pF_q(a_1,\dots,a_p;b_1,\dots,b_q;z)$ being the generalized hypergeometric function. Naturally, due to the intricacies of the hypergeometric function, an analytical solution for the gap equation above is not known. Thus, we proceed to a numerical analysis, which indicates that, for a given value of $M$, the gap equation will only admit solutions for $g > g_0$, similarly to what we have obtained in the previous case. The treatment of the gap equation in this scenario was first studied in~\cite{Felix:2021eoq}, adopting a similar approach, but in that work, the authors have truncated the sum to the first order, having as a consequence a restriction in the parameter space. Here we consider the complete Taylor series, therefore extending their results without restricting the available parameter space.

Considering the results obtained up to this point, there is a clear indication that the presence of a massive parameter ($m_\phi$ or $M$) creates a restriction of the form $g > g_0>0$ 
for the possible values of the gauge coupling, allowing for a real solution of the gap equation. That is, if we have $m_\phi \neq 0$ or $M \neq 0$, not every value of the gauge coupling is consistent with finding a real solution for the gap equation, which implies a restriction in the parameter space. The numerical analysis of the gap equation clearly shows this, as one can see in Figs.~\ref{M0} and~\ref{m_phi0}, where we considered the cases $M=0$ and $m_\phi=0$, respectively. Indeed, for a fixed value of $g^2$, a real solution for the gap equation can only be obtained below a given threshold in the mass parameters. This can be seen analytically from the equation~\eqref{condM0}, but it is hard to extract such a  conclusion from expression~\eqref{gapeqm0}.


\subsection{Case 4: $m_\phi \neq 0$ and $M \neq 0$}

\begin{figure}
    \centering
    \includegraphics[width=\linewidth]{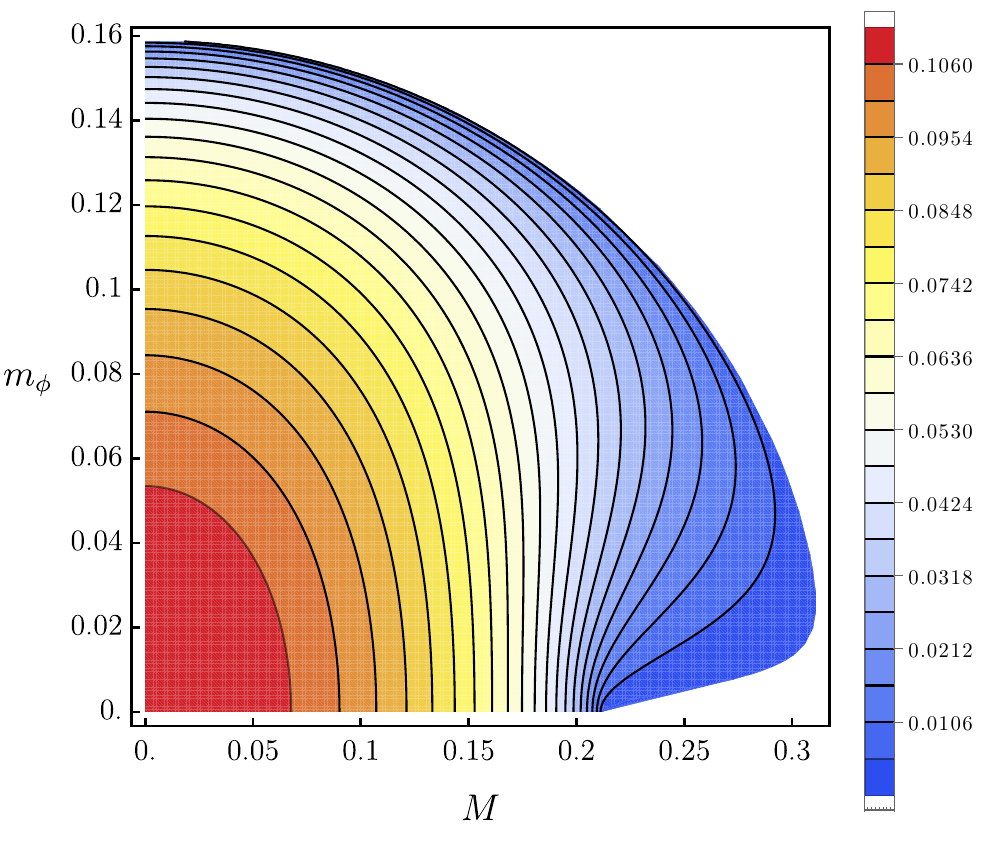}
    \caption{Contour plot of $\lambda$ as a function of $M$ and $m_\phi$, with a fixed value of $g^2=1.5$.}
    \label{contourplot}
\end{figure}

The gap equation in the general case where we have both $m_\phi \neq 0$ and $M \neq 0$ is more involved. We were unable to find analytical solutions in this case, which led us to perform a numerical analysis from the beginning. The procedure adopted was as follows: for a fixed value of the gauge coupling $g$ and of the mass parameters $M$ and $m_\phi$, we numerically solve the integral~\eqref{gapeq} to obtain $\lambda$. Then, keeping $g$ fixed, we vary the masses $M$ and $m_\phi$ in a suitable range. For each pair $\left(M, m_\phi\right)$, we find the corresponding $\lambda = \lambda\left(M, m_\phi\right)$ by solving the gap equation, which we plot to find out how the dependence of $\lambda$ in these masses is. For example, fixing $g^2=1.5$, we obtain Fig.~\ref{contourplot}. In this analysis, for definiteness, we set to zero any value $\lambda < 0.001$, to exclude any non-vanishing value produced by numerical errors. Naturally, there is some arbitrariness in the choice of this threshold that was chosen to be roughly two orders of magnitude below the value $\lambda(0,0) \simeq 0.113$.

The procedure described above allows us to constrain the parameter space according to the gap equation, which uniquely specifies $\lambda$ for given $g$ and $\left(M, m_\phi\right)$. In the next section, we restrict ourselves to the parameter space region that is not forbidden by the gap equation. 

Before moving to the analysis of the propagator, we make a brief comment on the behavior of $\lambda$ which can be observed in Fig.~\ref{contourplot}. Contrary to the cases 2 and 3, where only one mass parameter is included, in the presence of both $M$ and $m_\phi$, the Gribov parameter develops a non-monotonic dependence on these mass parameters. This seems to signal a competition between them on their contribution to $\lambda$. 

Although a complete explanation would require an analytical solution for the gap equation (which is beyond the purpose of this work), we can get a hint by looking at the derivatives of the left-hand side of the gap equation~\eqref{gapeq} with respect to these mass parameters, which are given by
\begin{align}\label{devM}
        &\frac{\partial}{\partial M^2}\int \frac{{\rm d}^3k}{(2\pi)^3}\frac{k^4 + m_\phi^2k^2 + \lambda^4}{(k^4 + m_\phi^2k^2 + \lambda^4 )^2 + M^2k^6} = \nonumber \\
        &=-\int \frac{{\rm d}^3k}{(2\pi)^3}\frac{k^8 (k^4 + m_\phi^2k^2  + \lambda^4)}{((k^4 + m_\phi^2k^2 + \lambda^4)^2+ M^2 k^6 )^2},
\end{align}
and
\begin{align}\label{devmphi}
        &\frac{\partial}{\partial m_\phi^2}\int \frac{{\rm d}^3k}{(2\pi)^3}\frac{k^4 + m_\phi^2k^2 + \lambda^4}{(k^4 + m_\phi^2k^2 + \lambda^4 )^2 + M^2k^6} = \nonumber \\
        &=\int \frac{{\rm d}^3k}{(2\pi)^3} \frac{k^4 (M^2k^6  - (k^4+m_\phi^2k^2 + \lambda^4)^2}{(M^2k^6 + (k^4 +m_\phi^2k^2 + \lambda^4)^2)^2}.
\end{align}
From equation~\eqref{devM}, we can see that the gap equation is monotonically decreasing with respect to $M$. However, the derivative with respect to $m_\phi^2$~\eqref{devmphi} introduces a relative minus sign between the terms containing $M$ and $m_\phi$ in the numerator, signaling a change in the behavior of $\lambda$ according to the specific values of $M$ and $m_\phi$. This indicates that the Gribov parameter feels the contribution of each mass parameter differently.

Having analyzed the constraint provided by the gap equation in the parameter space under different circumstances, we now proceed to investigate the analytic structure of the gluon propagator considering the effects of Gribov copies and taking into account these findings.

\section{The infrared structure of the gluon propagator\label{sec:irprop}}

\subsection{General analysis}

\begin{figure}
    \centering
    \includegraphics[width=\linewidth]{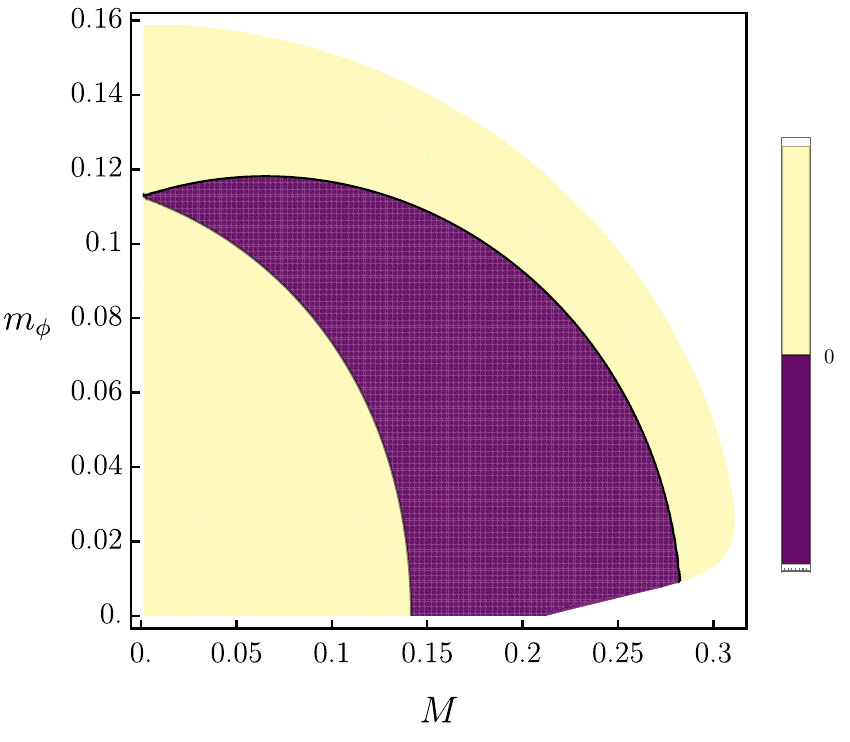}
    \caption{Sign of the discriminant $\Delta$ as a function of the parameters $M$ and $m_\phi$, with $\lambda$ defined through the gap equation~\eqref{gapeq} for a fixed value of $g^2=1.5$.}
    \label{disc}
\end{figure}

\begin{figure*}
  \includegraphics[width=0.48\linewidth]{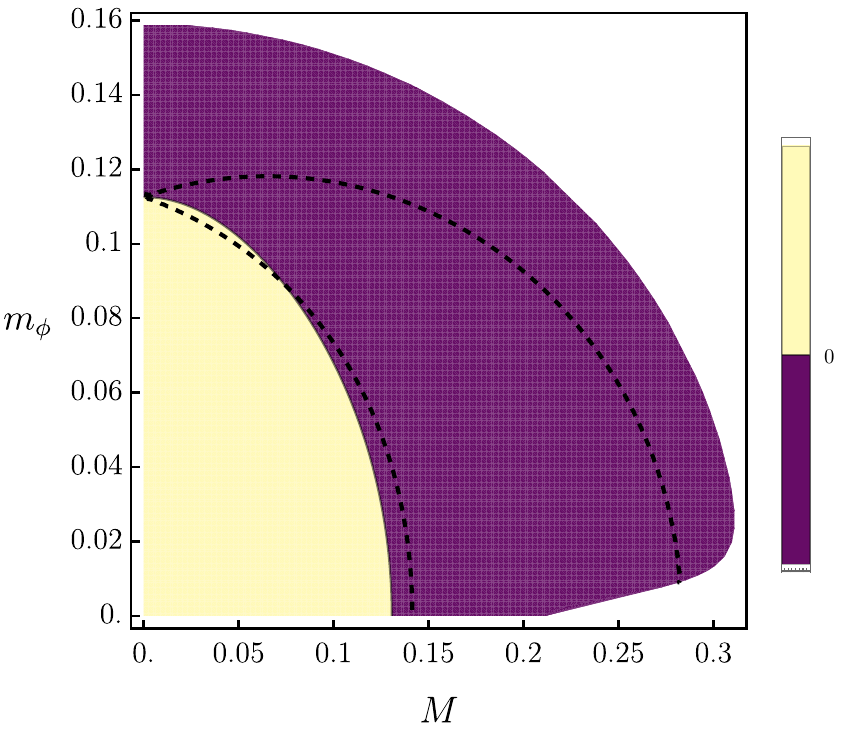}\hfill
  \includegraphics[width=0.48\linewidth]{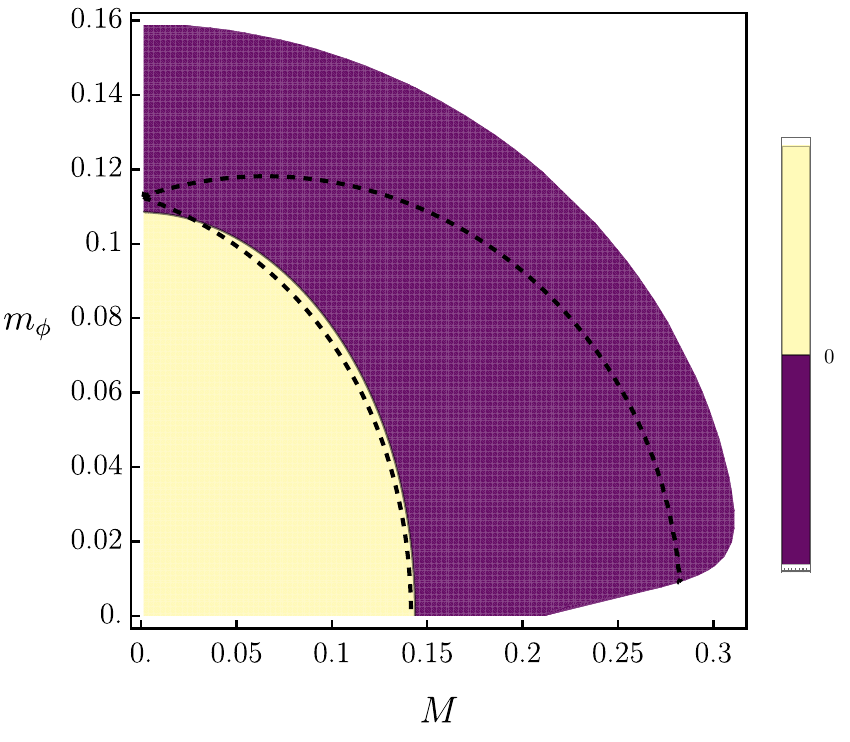}
  \caption{Sign of the subsidiary polynomials $P$ (left) and $D$ (right) as a function of the parameters $M$ and $m_\phi$, with $\lambda$ defined through the gap equation~\eqref{gapeq} for a fixed value of $g^2=1.5$.  The dashed curve indicates the change of sign of the discriminant $\Delta$.}
  \label{disc_sub}
\end{figure*}

The analytic structure of the gluon propagator gets modified in the infrared regime due to the existence of Gribov copies, as one can readily see in Eq.~\eqref{propgribov}. In particular, the underlying poles are changed and thus the corresponding spectrum is modified. The poles are the roots of $\mathcal{F}(k)$, defined as:
\begin{equation}\label{propnum}
	\mathcal{F}(k) = (k^4 + m_\phi^2k^2 + \lambda^4)^2 + M^2k^6\,,
\end{equation}
which can be conveniently rewritten as
\begin{equation}
	\mathcal{F}(k) = (k^2+m_1^2)(k^2+m_2^2)(k^2+m_3^2)(k^2+m_4^2)\,,
\end{equation}
for a certain set of parameters $m_i$, with $i=1,...,4$. In the Landau gauge ($\alpha \to 0$), the gluon propagator can be decomposed into parity-preserving and parity-violating contributions. That is,
\begin{align}
	\langle A^a_\mu(k) A^b_\nu(-k) \rangle = \delta^{ab}(\mathcal{K}_{\mu\nu}^{\text{pres}}(k) + \mathcal{K}_{\mu\nu}^{\text{viol}}(k))\,,
\end{align}
with
\begin{align}
	\mathcal{K}_{\mu\nu}^{\text{pres}}(k) &= \frac{(k^4 + m_\phi^2k^2 +\lambda^4)k^2}{(k^4 + m_\phi^2k^2 + \lambda^4)^2 + M^2k^6}\mathcal{P}^T_{\mu\nu}(k)\,, \nonumber \\
    \mathcal{K}_{\mu\nu}^{\text{viol}}(k) &= \frac{Mk^4}{(k^4 + m_\phi^2k^2 + \lambda^4)^2 + M^2k^6}\epsilon_{\mu\rho\nu}k_\rho\,.
\end{align}
    Next to that we can rewrite these terms using a partial fraction decomposition to make explicit their pole structure, {\it i.e.},
\begin{align}
	\mathcal{K}_{\mu\nu}^{\text{pres}} &= \sum_{i=1}^{4} \frac{A_i}{k^2+m_i^2}\mathcal{P}^T_{\mu\nu}(k), \nonumber \\
    \mathcal{K}_{\mu\nu}^{\text{viol}} &=\sum_{i=1}^{4} \frac{B_i}{k^2+m_i^2}\epsilon_{\mu\rho\nu}k_\rho,
\end{align}
where the factors $A_i$ and $B_i$ are expressed as
\begin{align}
	A_i &= \frac{m_i^2(m_i^4 - m_i^2 m_{\phi}^2 + \lambda^4)}{\Pi_{j=1,j\neq i}^{4}(m_i^2-m_j^2)}, \nonumber \\
    B_i &= -\frac{M m_i^4}{\Pi_{j=1,j\neq i}^{4}(m_i^2-m_j^2)}.
\end{align}

The gluon propagator expressed in this way is more suitable for analyzing its poles and residues, facilitating the discussion of the confining/non-confining nature of its excitations. The poles $m_i^2$ are functions of the CS mass $M$, the Higgs mass $m_\phi$, and the Gribov parameter $\lambda$. The gap equation~\eqref{gapeq} constrains these parameters, allowing us to find $\lambda$ for each given pair $\left(M,m_\phi\right)$, if we also specify the gauge coupling $g$. As already discussed, this imposes a constraint in the parameter space. 

Solving the gap equation to determine the allowed values of $(M,m_\phi,\lambda)$ is a distinctive feature of this work in comparison with previous studies~\cite{Canfora:2014qwe,Gomez:2016egs,Ferreira:2020hcj}, since in general these are considered as free parameters. Moreover, for the specific case where $m_\phi =0$, we considered the gap equation to all orders, instead of restricting it to the first order, as in~\cite{Felix:2021eoq}.

In the following, we use the results of Sec.~\ref{sec:gapeq} to analyze the analytic structure of the gluon propagator in the region of parameter space that is consistent with the gap equation, discussing where its excitations exhibit confining or non-confining behavior, characterized by the presence of complex or real poles, respectively. First of all, we rewrite the momentum as $k^2 \to x$ in Eq.~\eqref{propnum}, such that 
\begin{equation}
    x^4 + (M^2+2m_\phi^2) x^3+ (m_\phi^4+2\lambda^4) x^2+2m_\phi^2\lambda^4x+\lambda^8=0\,.
\end{equation}

The pole structure depends on the sign of the discriminant $\Delta$ of this quartic equation. For $\Delta>0$, the roots are either all real or all complex; for $\Delta<0$, the quartic polynomial has two distinct real roots and two complex conjugate roots. The discriminant is given by 
\begin{align}\label{discriminant}
    \Delta &= 16 \lambda ^{12} m_\phi^8 M^4+4 \lambda ^{12} m_\phi^6 M^6-128 \lambda ^{16} m_\phi^4 M^4 \nonumber \\
    &-144 \lambda ^{16} m_\phi^2 M^6-27 \lambda ^{16} M^8+256 \lambda ^{20} M^4\,.
\end{align}
The more relevant information here is given by the sign of the discriminant~\eqref{discriminant}. Considering only the region in parameter space consistent with the gap equation, we exhibit in Fig.~\ref{disc} the sign of the discriminant $\Delta$ for given values $\left(M, m_\phi\right)$, with the Gribov parameter determined by the gap equation for a fixed value of the gauge coupling (here chosen as $g^2=1.5$, for concreteness). As one can readily see in Fig.~\ref{disc}, the discriminant is positive in a significant part of the parameter space, indicating that we have either all roots being real or all of them coming in complex-conjugate pairs. To inspect which of these two cases occurs for each pair $(M,m_\phi)$, we have to analyze the subsidiary polynomials:
\begin{equation}\label{p}
	P(M,m_\phi,\lambda) = 8 (m_\phi^4 + 2 \lambda^4) -3 (2 m_\phi^2 + M^2)^2\,,
\end{equation}
and
\begin{align}\label{d}
    	D(M,m_\phi,\lambda) &= -M^2 \left(32 m_\phi^6 + 56 m_\phi^4 M^2 + 3 M^6 \right. \nonumber \\
        &\left. - 32 M^2 \lambda^4 + 24 m_\phi^2 (M^4 - 4 \lambda^4) \right)\,.
\end{align}

The sign of the subsidiary polynomials $P(M,m_\phi,\lambda)$ and $D(M,m_\phi,\lambda)$ is what determines whether the roots of $\mathcal{F}(k)$ are all real or complex-conjugate, for $\Delta>0$. For $P(M,m_\phi,\lambda)<0$ and $D(M,m_\phi,\lambda)<0$ they are all real; for $P(M,m_\phi,\lambda)>0$ or $D(M,m_\phi,\lambda)>0$, they are all complex-conjugate. We exhibit the signs of $P(M,m_\phi,\lambda)$ and $D(M,m_\phi,\lambda)$ for the relevant region of parameter space in Fig.~\ref{disc_sub}. 

\begin{figure}
    \centering
    \includegraphics[width=\linewidth]{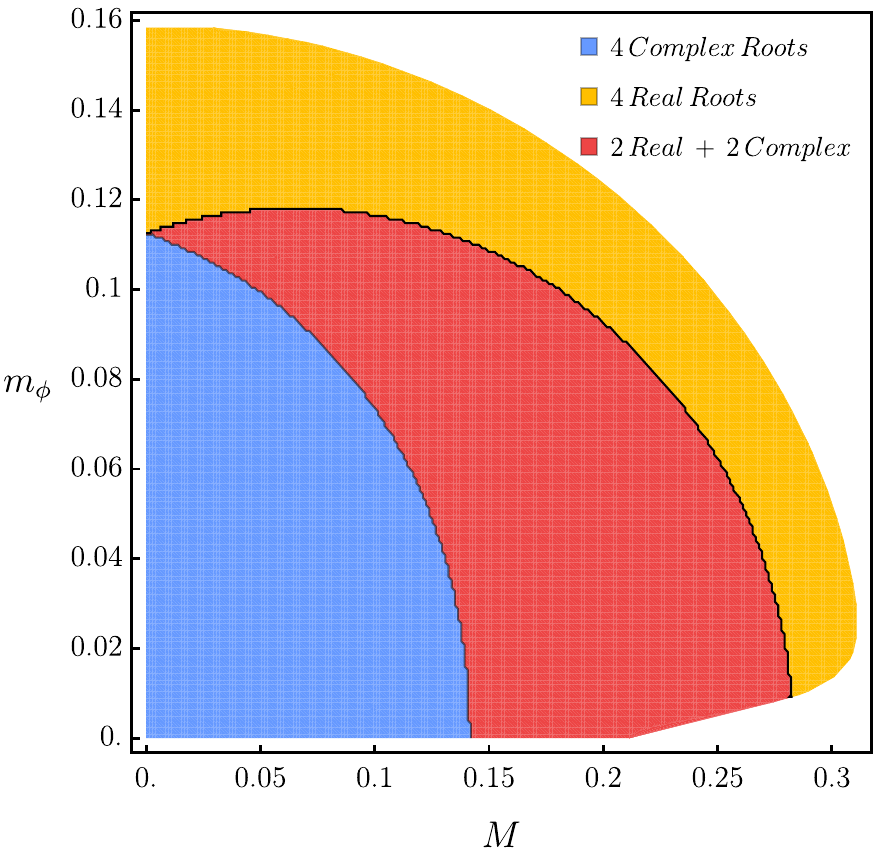}
    \caption{Change of the roots of the polynomial $\mathcal{F}(k)$ in terms of the mass parameters $M$ and $m_\phi$.}
    \label{RealRootsPlot}
\end{figure}

One can see that there is a region in parameter space where $\Delta>0$ with $P<0$ and $D<0$ outside the outer dotted line, which means that all roots are real. This provides candidates for physical excitations. Furthermore, there is also a region between the dotted lines where $\Delta<0$, leading to two real roots. Thus, it potentially accommodates physical excitations, provided that the associated residues are positive. On the other hand, there is a region inside the inner dotted line where $\Delta>0$ but $P>0$ or $D>0$\footnote{Even though there are slices in the parameter space where the subsidiary polynomials $P$ and $D$ are negative in this region, these slices do not overlap. Consequently, the condition $P<0$ and $D<0$, required for four real roots, is never satisfied, and all four roots remain complex.}, indicating that all roots are complex. The summary can be found in Fig.~\ref{RealRootsPlot}, where we can explicitly see the regions in which we have zero, two, and four real roots. 

From Fig.~\ref{RealRootsPlot}, one can readily see that a change in the spectrum takes place as we increase the mass parameters $m_\phi$ and/or $M$ (for a fixed value of the gauge coupling and consistently determining $\lambda$ by the gap equation). Indeed, we can start from a region characterized by small $m_\phi$ and $M$, in which there are no real roots. As the values of the mass parameters increase, we move to a region where there are two real roots. Finally, for larger values of $M$ and/or $m_\phi$, we can reach a region in which all four roots are real.

We can interpret the appearance of the real roots as we increase the mass parameters (while still obeying the consistency constraint of the gap equation) as signaling a departure from a totally confining phase, where all the poles are complex and therefore do not represent any physical observable excitation, passing through a mixed phase where there are both confined and deconfined excitations in the spectrum, and finally reaching a totally deconfined phase where all the poles are real.    

\begin{figure}
    \centering
    \includegraphics[width=\linewidth]{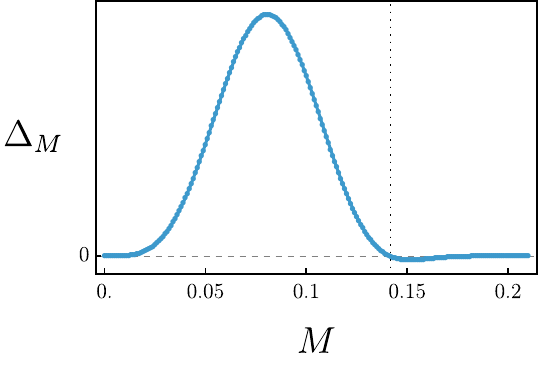}
    \caption{Discriminant $\Delta_M$ as a function of $M$, with $\lambda$ defined through the gap equation (\ref{gapeqM}) and a fixed value of $g^2=1.5$. The dotted vertical line indicates the point where the determinant  $\Delta_M$ changes sign. }
    \label{discM}
\end{figure}

\begin{figure*}
  \includegraphics[width=0.48\linewidth]{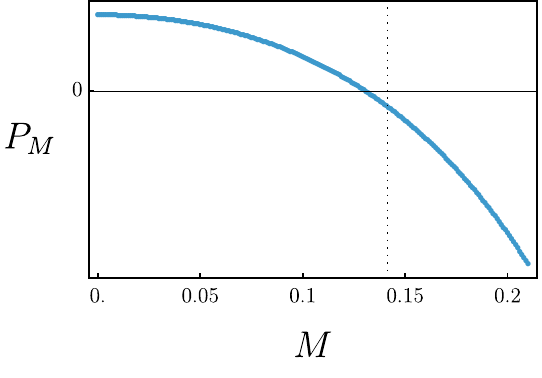}\hfill
  \includegraphics[width=0.48\linewidth]{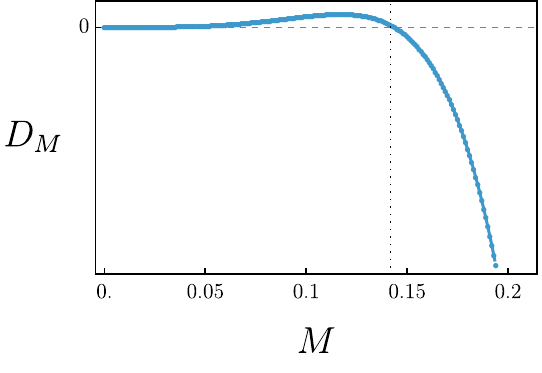}
  \caption{Subsidiary polynomials $P_M$ (left) and $D_M$ (right) as a function of $M$, with $\lambda$ defined through the gap equation (\ref{gapeqM}) and a fixed value of $g^2=1.5$. The dotted vertical line indicates the point where the determinant $\Delta_M$ changes sign.}
  \label{PM}
\end{figure*}

\subsection{The YMCS limit ($m_\phi=0$ and $M\neq0$)}

Let us consider now the analytic structure of the gluon propagator in the YMCS limit, that is, in the particular case where $m_\phi=0$, to see if we can get some insight into the role played by the Higgs sector in this analysis.

Naturally, in this particular case, the discriminant and the subsidiary polynomials have far simpler expressions:
\begin{equation}\label{discriminantM}
    \Delta_M= 256 \lambda ^{20} M^4 -27 \lambda ^{16} M^8\,,
\end{equation}
\begin{equation}\label{pM}
	P_M(M,\lambda) = 16 \lambda^4 -3 M^4\,,
\end{equation}
and
\begin{equation}\label{dM}
	D_M(M,\lambda) = 32 M^4 \lambda^4  -3 M^8\,.
\end{equation}
As before, the Gribov parameter $\lambda$ is determined (for a fixed value of the gauge coupling) through its gap equation for each value of $M$. The discriminant is shown in Fig.~\ref{discM}, where one can see that it has positive values up to $M\approx 0.14$, being negative for larger values of $M$. This is expected, because as we have shown before (see Fig.~\ref{m_phi0}), $\lambda$ decreases as $M$ increases (for fixed $g$) thanks to the gap equation constraint. Thus, as $M$ increases, the second term of Eq.~\eqref{discriminantM} dominates over the first one at some point, changing the sign of $\Delta_M$.

From the discussion presented above, we see that for $M>0.14$, $\Delta_M < 0$, thus implying two real roots. For the region where $\Delta_M>0$, we need to look at the subsidiary polynomials to whether the roots are real or complex. However, as can be seen from Fig.~\ref{PM}, for all values of $M$ which make $\Delta_M>0$, we have $P_M(M,\lambda)$ or $D_M(M,\lambda)$ also positive, leading us to conclude that all roots are complex in this case.

Therefore, we can conclude that the addition of a Higgs sector to the YMCS theory makes the analytic structure of the gluon propagator more intricate, allowing for a richer phase structure. Indeed, in the YMCSH theory, there are not only regions with zero and two real roots, but also a region in which all the roots are real.

\begin{figure}[t]
	\includegraphics[width=\linewidth]{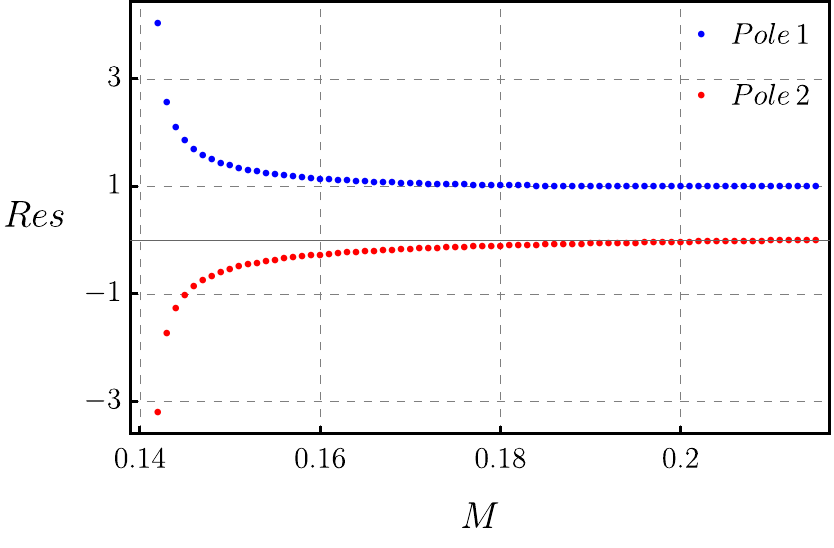}
	\caption{Residue of the parity preserving term of the propagator for the real roots as a function of $M$, with $\lambda$ defined through the gap equation (\ref{gapeqM}) and a fixed value of $g^2=1.5$. Note that we only have real roots beginning at $M\approx 0.14$, as shown by the determinant.}
	\label{residueM}
\end{figure}

To understand if the real roots stand a chance to describe physical excitations of the theory, one has to look at the residues at those poles to check if they are positive, since poles with negative residues would imply the presence of ghosts. From Fig.~\ref{discM}, we have that the two real roots appear in the interval $0.14\leq M \leq 0.21$, where in the upper limit the Gribov parameter goes to zero. Therefore, one of the roots must have positive residue in this interval for a physical excitation to exist. Moreover, since $\lambda \to 0$ as $M \to 0.21$, we recover the UV result in this limit, so the residue of the parity preserving term must be equal to 1 . Also, as we recover the UV expression for the propagator, the residue associated with the second root must vanish, since in this limit we have $\langle AA\rangle_{UV}^{\text{pres}}=(k^2+M^2)^{-1}\mathcal{P}^T_{\mu\nu}(k)$, meaning that this second excitation is not present. This behavior can be seen in Fig.~\ref{residueM}, and it works as a consistency check for the numerical solutions obtained here. Therefore, we have two massive real poles, one with a positive residue and the other with a negative residue, which is consistent with the UV limit. 

\subsection{The YM-Higgs limit ($m_\phi \neq 0$ and $M=0$)}

In the YM-Higgs limit, the gluon propagator is
\begin{align}\label{propgribovYMH}
		\langle A^a_\mu(k) A^b_\nu(-k) \rangle_{\rm H} &=  \frac{k^2}{(k^4 + m_\phi^2k^2 + \lambda^4)} \delta^{ab} \mathcal{P}^T_{\mu\nu}(k)\,.
\end{align}
In this case, the analysis is much simpler than in the previous ones, since we now have only a quadratic polynomial in $k^2$. The discriminant is simply $\Delta_{m_\phi}=m_\phi^4-4\lambda^4$, which has two distinct real roots for $\Delta_{m_\phi}>0$, a double real root for $\Delta_{m_\phi}=0$ and complex conjugate roots for $\Delta_{m_\phi}<0$. Taking into account the definition of the Gribov parameter~\eqref{gapeqYMH} and the restriction it imposes on the Higgs mass, we have complex roots for $g^4>2(9\pi^2m_\phi^2)$ and real roots for $9\pi^2m_\phi^2<g^4\leq2(9\pi^2m_\phi^2)$. In the limit $g^4=9\pi^2m_\phi^2$, the Gribov parameter goes to zero and we recover the UV expression for the propagator. For completeness, we mention that this is consistent with the results reported in~\cite{Capri:2012cr}, to where we refer the reader interested in a more detailed discussion about the physical excitations of this system.  

\section{Conclusions} \label{sec:conc}

In this work, we investigated the analytic structure of the gluon propagator in a three-dimensional Euclidean YMCS theory coupled to a Higgs field in the fundamental representation of the $SU(2)$ gauge group. We started analyzing the effect of eliminating infinitesimal Gribov copies by restricting the path integral to the Gribov region. This dynamically generates a mass scale associated with the Gribov horizon, changing the pole structure of the gluon propagator, which depends on three distinct mass scales, namely: the topological mass $M$ coming from the CS term; the scalar mass $m_\phi$ arising from the Higgs mechanism; the Gribov mass $\lambda$ related to the elimination of Gribov copies. 
Furthermore, the gap equation, which consistently determines the Gribov mass, can be explicitly solved in three dimensions, giving a constraint in parameter space that allows us to investigate the different phases of the theory by studying the interplay between the other two massive parameters. This provided a rich environment for understanding how the self-consistent determination of the Gribov parameter can affect the spectrum of the theory in contrast to the four-dimensional case where the solution of the gap equation is much more intricate.

The gap equation analysis revealed that there are important constraints in the parameter space that must be taken into account. This can be clearly seen in the limiting cases that we considered, which can be compared to previous works: i) if $m_\phi = M = 0$, then $\lambda$ is proportional to $g^2$~\eqref{GapEqCase1}; ii) if $M=0$ and $m_\phi \neq 0$, then we find a simple equation giving $\lambda$ in terms of $m_\phi$ and $g^2$~\eqref{gapeqYMH}; iii) if $m_\phi=0$ and $M \neq 0$, then we can relate $\lambda$, $M$, and $g^2$ in terms of generalized hypergeometric functions~\eqref{gapeqm0}. In the general case, the restriction imposed on $\lambda$ by the gap equation for a fixed value of $g^2$ can be seen as a function of $M$ and $m_\phi$ in the contour plot shown in Fig.~\ref{contourplot}.

Armed with the information coming from the gap equation, we investigated the analytic structure of the gluon propagator, discussing the different phases in which the theory can be, depending on the values of $M$ and $m_\phi$, for a fixed value of the coupling $g^2$. We have found regions in parameter space where all poles are real, indicating a deconfined phase with physical excitations of the fields, and have also found regions where complex poles are present, signaling a confined phase where there are excitations that are not present in the physical spectrum. 

In the particular case of YMCS ($m_\phi=0$), the phase structure was even simpler, divided into only two regions: for small values of $M$, the theory is confining, since all poles are complex; for larger values of $M$, there are two real poles, one with a positive residue for the parity-preserving sector, being a genuine physical excitation. The other has a negative residue, being a ghost. Moreover, in the limit $\lambda \to 0$, the positive residue goes to one and the negative residue goes to zero, recovering the standard perturbative behavior of YMCS.

A powerful tool for investigating the IR regime of YM theories is provided by lattice gauge theory. It would be highly desirable to compare our findings with lattice simulations, but the presence of a CS term makes the lattice implementation of such a system challenging. However, there have been many advances in related topics recently~\cite{Zhang:2024sgm,Xu:2024hyo,Peng:2025nfa}, giving us hope that the results presented here could be compared with lattice data in the future.  

Finally, as discussed in \cite{Ferreira:2021ksd}, the removal of infinitesimal Gribov copies in the path integral of YMCS theories quantized in linear covariant gauges seems to suffer from IR instabilities that lead to the formation of condensates, see \cite{Dudal:2008sp}. Thus, a refined version of YMCSH-GZ action can be constructed introducing new mass parameters that are not free but dynamically fixed by their own gap equations. These extra parameters will affect the discussion of the spectrum of theory. A comprehensive investigation of this is left for future work.

\section*{Acknowledgments}
DORA acknowledges the PEDECIBA program and the ANII-FCE-2025-186497 project for financial support.
PDF acknowledges the National Council for Scientific and Technological Development – CNPq for the financial support (No. 402459/2024-5).
GPB acknowledges CNPq under the grants PQ-C 308651/2025-1 and 406997/2025-0.
ADP acknowledges CNPq under the grant PQ-2 (312211/2022-8), FAPERJ under the “Jovem Cientista do Nosso Estado” program (E-26/204.457/2025).

\bibliography{refs}

\begin{thebibliography}{57}%
\makeatletter
\providecommand \@ifxundefined [1]{%
 \@ifx{#1\undefined}
}%
\providecommand \@ifnum [1]{%
 \ifnum #1\expandafter \@firstoftwo
 \else \expandafter \@secondoftwo
 \fi
}%
\providecommand \@ifx [1]{%
 \ifx #1\expandafter \@firstoftwo
 \else \expandafter \@secondoftwo
 \fi
}%
\providecommand \natexlab [1]{#1}%
\providecommand \enquote  [1]{``#1''}%
\providecommand \bibnamefont  [1]{#1}%
\providecommand \bibfnamefont [1]{#1}%
\providecommand \citenamefont [1]{#1}%
\providecommand \href@noop [0]{\@secondoftwo}%
\providecommand \href [0]{\begingroup \@sanitize@url \@href}%
\providecommand \@href[1]{\@@startlink{#1}\@@href}%
\providecommand \@@href[1]{\endgroup#1\@@endlink}%
\providecommand \@sanitize@url [0]{\catcode `\\12\catcode `\$12\catcode
  `\&12\catcode `\#12\catcode `\^12\catcode `\_12\catcode `\%12\relax}%
\providecommand \@@startlink[1]{}%
\providecommand \@@endlink[0]{}%
\providecommand \url  [0]{\begingroup\@sanitize@url \@url }%
\providecommand \@url [1]{\endgroup\@href {#1}{\urlprefix }}%
\providecommand \urlprefix  [0]{URL }%
\providecommand \Eprint [0]{\href }%
\providecommand \doibase [0]{https://doi.org/}%
\providecommand \selectlanguage [0]{\@gobble}%
\providecommand \bibinfo  [0]{\@secondoftwo}%
\providecommand \bibfield  [0]{\@secondoftwo}%
\providecommand \translation [1]{[#1]}%
\providecommand \BibitemOpen [0]{}%
\providecommand \bibitemStop [0]{}%
\providecommand \bibitemNoStop [0]{.\EOS\space}%
\providecommand \EOS [0]{\spacefactor3000\relax}%
\providecommand \BibitemShut  [1]{\csname bibitem#1\endcsname}%
\let\auto@bib@innerbib\@empty
\bibitem [{\citenamefont {Gross}\ and\ \citenamefont
  {Wilczek}(1973)}]{Gross:1973id}%
  \BibitemOpen
  \bibfield  {author} {\bibinfo {author} {\bibfnamefont {D.~J.}\ \bibnamefont
  {Gross}}\ and\ \bibinfo {author} {\bibfnamefont {F.}~\bibnamefont
  {Wilczek}},\ }\bibfield  {title} {\bibinfo {title} {{Ultraviolet Behavior of
  Nonabelian Gauge Theories}},\ }\href
  {https://doi.org/10.1103/PhysRevLett.30.1343} {\bibfield  {journal} {\bibinfo
   {journal} {Phys. Rev. Lett.}\ }\textbf {\bibinfo {volume} {30}},\ \bibinfo
  {pages} {1343} (\bibinfo {year} {1973})}\BibitemShut {NoStop}%
\bibitem [{\citenamefont {Politzer}(1973)}]{Politzer:1973fx}%
  \BibitemOpen
  \bibfield  {author} {\bibinfo {author} {\bibfnamefont {H.~D.}\ \bibnamefont
  {Politzer}},\ }\bibfield  {title} {\bibinfo {title} {{Reliable Perturbative
  Results for Strong Interactions?}},\ }\href
  {https://doi.org/10.1103/PhysRevLett.30.1346} {\bibfield  {journal} {\bibinfo
   {journal} {Phys. Rev. Lett.}\ }\textbf {\bibinfo {volume} {30}},\ \bibinfo
  {pages} {1346} (\bibinfo {year} {1973})}\BibitemShut {NoStop}%
\bibitem [{\citenamefont {Greensite}(2011)}]{Greensite:2011zz}%
  \BibitemOpen
  \bibfield  {author} {\bibinfo {author} {\bibfnamefont {J.}~\bibnamefont
  {Greensite}},\ }\href {https://doi.org/10.1007/978-3-642-14382-3} {\emph
  {\bibinfo {title} {{An introduction to the confinement problem}}}},\ Vol.\
  \bibinfo {volume} {821}\ (\bibinfo {year} {2011})\BibitemShut {NoStop}%
\bibitem [{\citenamefont {Brambilla}\ \emph {et~al.}(2014)\citenamefont
  {Brambilla} \emph {et~al.}}]{Brambilla:2014jmp}%
  \BibitemOpen
  \bibfield  {author} {\bibinfo {author} {\bibfnamefont {N.}~\bibnamefont
  {Brambilla}} \emph {et~al.},\ }\bibfield  {title} {\bibinfo {title} {{QCD and
  Strongly Coupled Gauge Theories: Challenges and Perspectives}},\ }\href
  {https://doi.org/10.1140/epjc/s10052-014-2981-5} {\bibfield  {journal}
  {\bibinfo  {journal} {Eur. Phys. J. C}\ }\textbf {\bibinfo {volume} {74}},\
  \bibinfo {pages} {2981} (\bibinfo {year} {2014})},\ \Eprint
  {https://arxiv.org/abs/1404.3723} {arXiv:1404.3723 [hep-ph]} \BibitemShut
  {NoStop}%
\bibitem [{\citenamefont {Faddeev}\ and\ \citenamefont
  {Popov}(1967)}]{Faddeev:1967fc}%
  \BibitemOpen
  \bibfield  {author} {\bibinfo {author} {\bibfnamefont {L.}~\bibnamefont
  {Faddeev}}\ and\ \bibinfo {author} {\bibfnamefont {V.}~\bibnamefont
  {Popov}},\ }\bibfield  {title} {\bibinfo {title} {{Feynman Diagrams for the
  Yang-Mills Field}},\ }\href {https://doi.org/10.1016/0370-2693(67)90067-6}
  {\bibfield  {journal} {\bibinfo  {journal} {Phys. Lett. B}\ }\textbf
  {\bibinfo {volume} {25}},\ \bibinfo {pages} {29} (\bibinfo {year}
  {1967})}\BibitemShut {NoStop}%
\bibitem [{\citenamefont {Gribov}(1978)}]{Gribov:1977wm}%
  \BibitemOpen
  \bibfield  {author} {\bibinfo {author} {\bibfnamefont {V.~N.}\ \bibnamefont
  {Gribov}},\ }\bibfield  {title} {\bibinfo {title} {{Quantization of
  Nonabelian Gauge Theories}},\ }\href
  {https://doi.org/10.1016/0550-3213(78)90175-X} {\bibfield  {journal}
  {\bibinfo  {journal} {Nucl. Phys.}\ }\textbf {\bibinfo {volume} {B139}},\
  \bibinfo {pages} {1} (\bibinfo {year} {1978})},\ \bibinfo {note}
  {[1(1977)]}\BibitemShut {NoStop}%
\bibitem [{\citenamefont {Singer}(1978)}]{Singer:1978dk}%
  \BibitemOpen
  \bibfield  {author} {\bibinfo {author} {\bibfnamefont {I.}~\bibnamefont
  {Singer}},\ }\bibfield  {title} {\bibinfo {title} {{Some Remarks on the
  Gribov Ambiguity}},\ }\href {https://doi.org/10.1007/BF01609471} {\bibfield
  {journal} {\bibinfo  {journal} {Commun. Math. Phys.}\ }\textbf {\bibinfo
  {volume} {60}},\ \bibinfo {pages} {7} (\bibinfo {year} {1978})}\BibitemShut
  {NoStop}%
\bibitem [{\citenamefont {Vandersickel}\ and\ \citenamefont
  {Zwanziger}(2012)}]{Vandersickel:2012tz}%
  \BibitemOpen
  \bibfield  {author} {\bibinfo {author} {\bibfnamefont {N.}~\bibnamefont
  {Vandersickel}}\ and\ \bibinfo {author} {\bibfnamefont {D.}~\bibnamefont
  {Zwanziger}},\ }\bibfield  {title} {\bibinfo {title} {{The Gribov problem and
  QCD dynamics}},\ }\href {https://doi.org/10.1016/j.physrep.2012.07.003}
  {\bibfield  {journal} {\bibinfo  {journal} {Phys. Rept.}\ }\textbf {\bibinfo
  {volume} {520}},\ \bibinfo {pages} {175} (\bibinfo {year} {2012})},\ \Eprint
  {https://arxiv.org/abs/1202.1491} {arXiv:1202.1491 [hep-th]} \BibitemShut
  {NoStop}%
\bibitem [{\citenamefont {Sobreiro}\ and\ \citenamefont
  {Sorella}(2005)}]{Sobreiro:2005ec}%
  \BibitemOpen
  \bibfield  {author} {\bibinfo {author} {\bibfnamefont {R.}~\bibnamefont
  {Sobreiro}}\ and\ \bibinfo {author} {\bibfnamefont {S.}~\bibnamefont
  {Sorella}},\ }\bibfield  {title} {\bibinfo {title} {{Introduction to the
  Gribov ambiguities in Euclidean Yang-Mills theories}},\ }in\ \href@noop {}
  {\emph {\bibinfo {booktitle} {{13th Jorge Andre Swieca Summer School on
  Particle and Fields}}}}\ (\bibinfo {year} {2005})\ \Eprint
  {https://arxiv.org/abs/hep-th/0504095} {arXiv:hep-th/0504095} \BibitemShut
  {NoStop}%
\bibitem [{\citenamefont {Zwanziger}(1989{\natexlab{a}})}]{Zwanziger:1989mf}%
  \BibitemOpen
  \bibfield  {author} {\bibinfo {author} {\bibfnamefont {D.}~\bibnamefont
  {Zwanziger}},\ }\bibfield  {title} {\bibinfo {title} {{Local and
  Renormalizable Action From the Gribov Horizon}},\ }\href
  {https://doi.org/10.1016/0550-3213(89)90122-3} {\bibfield  {journal}
  {\bibinfo  {journal} {Nucl. Phys.}\ }\textbf {\bibinfo {volume} {B323}},\
  \bibinfo {pages} {513} (\bibinfo {year} {1989}{\natexlab{a}})}\BibitemShut
  {NoStop}%
\bibitem [{\citenamefont {Dudal}\ \emph
  {et~al.}(2008{\natexlab{a}})\citenamefont {Dudal}, \citenamefont {Sorella},
  \citenamefont {Vandersickel},\ and\ \citenamefont
  {Verschelde}}]{Dudal:2007cw}%
  \BibitemOpen
  \bibfield  {author} {\bibinfo {author} {\bibfnamefont {D.}~\bibnamefont
  {Dudal}}, \bibinfo {author} {\bibfnamefont {S.~P.}\ \bibnamefont {Sorella}},
  \bibinfo {author} {\bibfnamefont {N.}~\bibnamefont {Vandersickel}},\ and\
  \bibinfo {author} {\bibfnamefont {H.}~\bibnamefont {Verschelde}},\ }\bibfield
   {title} {\bibinfo {title} {{New features of the gluon and ghost propagator
  in the infrared region from the Gribov-Zwanziger approach}},\ }\href
  {https://doi.org/10.1103/PhysRevD.77.071501} {\bibfield  {journal} {\bibinfo
  {journal} {Phys. Rev.}\ }\textbf {\bibinfo {volume} {D77}},\ \bibinfo {pages}
  {071501} (\bibinfo {year} {2008}{\natexlab{a}})},\ \Eprint
  {https://arxiv.org/abs/0711.4496} {arXiv:0711.4496 [hep-th]} \BibitemShut
  {NoStop}%
\bibitem [{\citenamefont {Dudal}\ \emph
  {et~al.}(2008{\natexlab{b}})\citenamefont {Dudal}, \citenamefont {Gracey},
  \citenamefont {Sorella}, \citenamefont {Vandersickel},\ and\ \citenamefont
  {Verschelde}}]{Dudal:2008sp}%
  \BibitemOpen
  \bibfield  {author} {\bibinfo {author} {\bibfnamefont {D.}~\bibnamefont
  {Dudal}}, \bibinfo {author} {\bibfnamefont {J.~A.}\ \bibnamefont {Gracey}},
  \bibinfo {author} {\bibfnamefont {S.~P.}\ \bibnamefont {Sorella}}, \bibinfo
  {author} {\bibfnamefont {N.}~\bibnamefont {Vandersickel}},\ and\ \bibinfo
  {author} {\bibfnamefont {H.}~\bibnamefont {Verschelde}},\ }\bibfield  {title}
  {\bibinfo {title} {{A Refinement of the Gribov-Zwanziger approach in the
  Landau gauge: Infrared propagators in harmony with the lattice results}},\
  }\href {https://doi.org/10.1103/PhysRevD.78.065047} {\bibfield  {journal}
  {\bibinfo  {journal} {Phys. Rev.}\ }\textbf {\bibinfo {volume} {D78}},\
  \bibinfo {pages} {065047} (\bibinfo {year} {2008}{\natexlab{b}})},\ \Eprint
  {https://arxiv.org/abs/0806.4348} {arXiv:0806.4348 [hep-th]} \BibitemShut
  {NoStop}%
\bibitem [{\citenamefont {Dudal}\ \emph {et~al.}(2011)\citenamefont {Dudal},
  \citenamefont {Sorella},\ and\ \citenamefont {Vandersickel}}]{Dudal:2011gd}%
  \BibitemOpen
  \bibfield  {author} {\bibinfo {author} {\bibfnamefont {D.}~\bibnamefont
  {Dudal}}, \bibinfo {author} {\bibfnamefont {S.~P.}\ \bibnamefont {Sorella}},\
  and\ \bibinfo {author} {\bibfnamefont {N.}~\bibnamefont {Vandersickel}},\
  }\bibfield  {title} {\bibinfo {title} {{The dynamical origin of the
  refinement of the Gribov-Zwanziger theory}},\ }\href
  {https://doi.org/10.1103/PhysRevD.84.065039} {\bibfield  {journal} {\bibinfo
  {journal} {Phys. Rev. D}\ }\textbf {\bibinfo {volume} {84}},\ \bibinfo
  {pages} {065039} (\bibinfo {year} {2011})},\ \Eprint
  {https://arxiv.org/abs/1105.3371} {arXiv:1105.3371 [hep-th]} \BibitemShut
  {NoStop}%
\bibitem [{\citenamefont {van Baal}(1992)}]{vanBaal:1991zw}%
  \BibitemOpen
  \bibfield  {author} {\bibinfo {author} {\bibfnamefont {P.}~\bibnamefont {van
  Baal}},\ }\bibfield  {title} {\bibinfo {title} {{More (thoughts on) Gribov
  copies}},\ }\href {https://doi.org/10.1016/0550-3213(92)90386-P} {\bibfield
  {journal} {\bibinfo  {journal} {Nucl. Phys.}\ }\textbf {\bibinfo {volume}
  {B369}},\ \bibinfo {pages} {259} (\bibinfo {year} {1992})}\BibitemShut
  {NoStop}%
\bibitem [{\citenamefont {Dell'Antonio}\ and\ \citenamefont
  {Zwanziger}(1991)}]{DellAntonio:1991mms}%
  \BibitemOpen
  \bibfield  {author} {\bibinfo {author} {\bibfnamefont {G.}~\bibnamefont
  {Dell'Antonio}}\ and\ \bibinfo {author} {\bibfnamefont {D.}~\bibnamefont
  {Zwanziger}},\ }\bibfield  {title} {\bibinfo {title} {{Every gauge orbit
  passes inside the Gribov horizon}},\ }\href
  {https://doi.org/10.1007/BF02099494} {\bibfield  {journal} {\bibinfo
  {journal} {Commun. Math. Phys.}\ }\textbf {\bibinfo {volume} {138}},\
  \bibinfo {pages} {291} (\bibinfo {year} {1991})}\BibitemShut {NoStop}%
\bibitem [{\citenamefont {Zwanziger}(1989{\natexlab{b}})}]{Zwanziger:1988jt}%
  \BibitemOpen
  \bibfield  {author} {\bibinfo {author} {\bibfnamefont {D.}~\bibnamefont
  {Zwanziger}},\ }\bibfield  {title} {\bibinfo {title} {{Action From the Gribov
  Horizon}},\ }\href {https://doi.org/10.1016/0550-3213(89)90263-0} {\bibfield
  {journal} {\bibinfo  {journal} {Nucl. Phys. B}\ }\textbf {\bibinfo {volume}
  {321}},\ \bibinfo {pages} {591} (\bibinfo {year}
  {1989}{\natexlab{b}})}\BibitemShut {NoStop}%
\bibitem [{\citenamefont {Sternbeck}\ \emph {et~al.}(2007)\citenamefont
  {Sternbeck}, \citenamefont {von Smekal}, \citenamefont {Leinweber},\ and\
  \citenamefont {Williams}}]{Sternbeck:2007ug}%
  \BibitemOpen
  \bibfield  {author} {\bibinfo {author} {\bibfnamefont {A.}~\bibnamefont
  {Sternbeck}}, \bibinfo {author} {\bibfnamefont {L.}~\bibnamefont {von
  Smekal}}, \bibinfo {author} {\bibfnamefont {D.~B.}\ \bibnamefont
  {Leinweber}},\ and\ \bibinfo {author} {\bibfnamefont {A.~G.}\ \bibnamefont
  {Williams}},\ }\bibfield  {title} {\bibinfo {title} {{Comparing SU(2) to
  SU(3) gluodynamics on large lattices}},\ }\href
  {https://doi.org/10.22323/1.042.0340} {\bibfield  {journal} {\bibinfo
  {journal} {PoS}\ }\textbf {\bibinfo {volume} {LATTICE2007}},\ \bibinfo
  {pages} {340} (\bibinfo {year} {2007})},\ \Eprint
  {https://arxiv.org/abs/0710.1982} {arXiv:0710.1982 [hep-lat]} \BibitemShut
  {NoStop}%
\bibitem [{\citenamefont {Cucchieri}\ and\ \citenamefont
  {Mendes}(2007)}]{Cucchieri:2007md}%
  \BibitemOpen
  \bibfield  {author} {\bibinfo {author} {\bibfnamefont {A.}~\bibnamefont
  {Cucchieri}}\ and\ \bibinfo {author} {\bibfnamefont {T.}~\bibnamefont
  {Mendes}},\ }\bibfield  {title} {\bibinfo {title} {{What's up with IR gluon
  and ghost propagators in Landau gauge? A puzzling answer from huge
  lattices}},\ }\href {https://doi.org/10.22323/1.042.0297} {\bibfield
  {journal} {\bibinfo  {journal} {PoS}\ }\textbf {\bibinfo {volume}
  {LATTICE2007}},\ \bibinfo {pages} {297} (\bibinfo {year} {2007})},\ \Eprint
  {https://arxiv.org/abs/0710.0412} {arXiv:0710.0412 [hep-lat]} \BibitemShut
  {NoStop}%
\bibitem [{\citenamefont {Cucchieri}\ and\ \citenamefont
  {Mendes}(2008)}]{Cucchieri:2007rg}%
  \BibitemOpen
  \bibfield  {author} {\bibinfo {author} {\bibfnamefont {A.}~\bibnamefont
  {Cucchieri}}\ and\ \bibinfo {author} {\bibfnamefont {T.}~\bibnamefont
  {Mendes}},\ }\bibfield  {title} {\bibinfo {title} {{Constraints on the IR
  behavior of the gluon propagator in Yang-Mills theories}},\ }\href
  {https://doi.org/10.1103/PhysRevLett.100.241601} {\bibfield  {journal}
  {\bibinfo  {journal} {Phys. Rev. Lett.}\ }\textbf {\bibinfo {volume} {100}},\
  \bibinfo {pages} {241601} (\bibinfo {year} {2008})},\ \Eprint
  {https://arxiv.org/abs/0712.3517} {arXiv:0712.3517 [hep-lat]} \BibitemShut
  {NoStop}%
\bibitem [{\citenamefont {Bornyakov}\ \emph {et~al.}(2009)\citenamefont
  {Bornyakov}, \citenamefont {Mitrjushkin},\ and\ \citenamefont
  {Muller-Preussker}}]{Bornyakov:2008yx}%
  \BibitemOpen
  \bibfield  {author} {\bibinfo {author} {\bibfnamefont {V.~G.}\ \bibnamefont
  {Bornyakov}}, \bibinfo {author} {\bibfnamefont {V.~K.}\ \bibnamefont
  {Mitrjushkin}},\ and\ \bibinfo {author} {\bibfnamefont {M.}~\bibnamefont
  {Muller-Preussker}},\ }\bibfield  {title} {\bibinfo {title} {{Infrared
  behavior and Gribov ambiguity in SU(2) lattice gauge theory}},\ }\href
  {https://doi.org/10.1103/PhysRevD.79.074504} {\bibfield  {journal} {\bibinfo
  {journal} {Phys. Rev. D}\ }\textbf {\bibinfo {volume} {79}},\ \bibinfo
  {pages} {074504} (\bibinfo {year} {2009})},\ \Eprint
  {https://arxiv.org/abs/0812.2761} {arXiv:0812.2761 [hep-lat]} \BibitemShut
  {NoStop}%
\bibitem [{\citenamefont {Bogolubsky}\ \emph {et~al.}(2009)\citenamefont
  {Bogolubsky}, \citenamefont {Ilgenfritz}, \citenamefont {Muller-Preussker},\
  and\ \citenamefont {Sternbeck}}]{Bogolubsky:2009dc}%
  \BibitemOpen
  \bibfield  {author} {\bibinfo {author} {\bibfnamefont {I.~L.}\ \bibnamefont
  {Bogolubsky}}, \bibinfo {author} {\bibfnamefont {E.~M.}\ \bibnamefont
  {Ilgenfritz}}, \bibinfo {author} {\bibfnamefont {M.}~\bibnamefont
  {Muller-Preussker}},\ and\ \bibinfo {author} {\bibfnamefont {A.}~\bibnamefont
  {Sternbeck}},\ }\bibfield  {title} {\bibinfo {title} {{Lattice gluodynamics
  computation of Landau gauge Green's functions in the deep infrared}},\ }\href
  {https://doi.org/10.1016/j.physletb.2009.04.076} {\bibfield  {journal}
  {\bibinfo  {journal} {Phys. Lett.}\ }\textbf {\bibinfo {volume} {B676}},\
  \bibinfo {pages} {69} (\bibinfo {year} {2009})},\ \Eprint
  {https://arxiv.org/abs/0901.0736} {arXiv:0901.0736 [hep-lat]} \BibitemShut
  {NoStop}%
\bibitem [{\citenamefont {Cucchieri}\ \emph {et~al.}(2012)\citenamefont
  {Cucchieri}, \citenamefont {Dudal}, \citenamefont {Mendes},\ and\
  \citenamefont {Vandersickel}}]{Cucchieri:2011ig}%
  \BibitemOpen
  \bibfield  {author} {\bibinfo {author} {\bibfnamefont {A.}~\bibnamefont
  {Cucchieri}}, \bibinfo {author} {\bibfnamefont {D.}~\bibnamefont {Dudal}},
  \bibinfo {author} {\bibfnamefont {T.}~\bibnamefont {Mendes}},\ and\ \bibinfo
  {author} {\bibfnamefont {N.}~\bibnamefont {Vandersickel}},\ }\bibfield
  {title} {\bibinfo {title} {{Modeling the Gluon Propagator in Landau Gauge:
  Lattice Estimates of Pole Masses and Dimension-Two Condensates}},\ }\href
  {https://doi.org/10.1103/PhysRevD.85.094513} {\bibfield  {journal} {\bibinfo
  {journal} {Phys. Rev. D}\ }\textbf {\bibinfo {volume} {85}},\ \bibinfo
  {pages} {094513} (\bibinfo {year} {2012})},\ \Eprint
  {https://arxiv.org/abs/1111.2327} {arXiv:1111.2327 [hep-lat]} \BibitemShut
  {NoStop}%
\bibitem [{\citenamefont {Cucchieri}\ \emph {et~al.}(2016)\citenamefont
  {Cucchieri}, \citenamefont {Dudal}, \citenamefont {Mendes},\ and\
  \citenamefont {Vandersickel}}]{Cucchieri:2016jwg}%
  \BibitemOpen
  \bibfield  {author} {\bibinfo {author} {\bibfnamefont {A.}~\bibnamefont
  {Cucchieri}}, \bibinfo {author} {\bibfnamefont {D.}~\bibnamefont {Dudal}},
  \bibinfo {author} {\bibfnamefont {T.}~\bibnamefont {Mendes}},\ and\ \bibinfo
  {author} {\bibfnamefont {N.}~\bibnamefont {Vandersickel}},\ }\bibfield
  {title} {\bibinfo {title} {{Modeling the Landau-gauge ghost propagator in 2,
  3, and 4 spacetime dimensions}},\ }\href
  {https://doi.org/10.1103/PhysRevD.93.094513} {\bibfield  {journal} {\bibinfo
  {journal} {Phys. Rev. D}\ }\textbf {\bibinfo {volume} {93}},\ \bibinfo
  {pages} {094513} (\bibinfo {year} {2016})},\ \Eprint
  {https://arxiv.org/abs/1602.01646} {arXiv:1602.01646 [hep-lat]} \BibitemShut
  {NoStop}%
\bibitem [{\citenamefont {de~Brito}\ and\ \citenamefont
  {Pereira}(2024)}]{deBrito:2024ffa}%
  \BibitemOpen
  \bibfield  {author} {\bibinfo {author} {\bibfnamefont {G.~P.}\ \bibnamefont
  {de~Brito}}\ and\ \bibinfo {author} {\bibfnamefont {A.~D.}\ \bibnamefont
  {Pereira}},\ }\bibfield  {title} {\bibinfo {title} {{Infrared gluon
  propagator in the refined Gribov-Zwanziger scenario at one-loop order in the
  Landau gauge}},\ }\href {https://doi.org/10.1103/PhysRevD.110.074005}
  {\bibfield  {journal} {\bibinfo  {journal} {Phys. Rev. D}\ }\textbf {\bibinfo
  {volume} {110}},\ \bibinfo {pages} {074005} (\bibinfo {year} {2024})},\
  \Eprint {https://arxiv.org/abs/2405.07779} {arXiv:2405.07779 [hep-th]}
  \BibitemShut {NoStop}%
\bibitem [{\citenamefont {de~Brito}\ \emph {et~al.}(2023)\citenamefont
  {de~Brito}, \citenamefont {De~Fabritiis},\ and\ \citenamefont
  {Pereira}}]{deBrito:2023qfs}%
  \BibitemOpen
  \bibfield  {author} {\bibinfo {author} {\bibfnamefont {G.~P.}\ \bibnamefont
  {de~Brito}}, \bibinfo {author} {\bibfnamefont {P.}~\bibnamefont
  {De~Fabritiis}},\ and\ \bibinfo {author} {\bibfnamefont {A.~D.}\ \bibnamefont
  {Pereira}},\ }\bibfield  {title} {\bibinfo {title} {{Refined Gribov-Zwanziger
  theory coupled to scalar fields in the Landau gauge}},\ }\href
  {https://doi.org/10.1103/PhysRevD.107.114006} {\bibfield  {journal} {\bibinfo
   {journal} {Phys. Rev. D}\ }\textbf {\bibinfo {volume} {107}},\ \bibinfo
  {pages} {114006} (\bibinfo {year} {2023})},\ \Eprint
  {https://arxiv.org/abs/2302.04827} {arXiv:2302.04827 [hep-th]} \BibitemShut
  {NoStop}%
\bibitem [{\citenamefont {de~Brito}\ \emph {et~al.}(2025)\citenamefont
  {de~Brito}, \citenamefont {De~Fabritiis},\ and\ \citenamefont
  {Pereira}}]{deBrito:2025nvl}%
  \BibitemOpen
  \bibfield  {author} {\bibinfo {author} {\bibfnamefont {G.~P.}\ \bibnamefont
  {de~Brito}}, \bibinfo {author} {\bibfnamefont {P.}~\bibnamefont
  {De~Fabritiis}},\ and\ \bibinfo {author} {\bibfnamefont {A.~D.}\ \bibnamefont
  {Pereira}},\ }\bibfield  {title} {\bibinfo {title} {{Quark propagator at one
  loop in the refined Gribov-Zwanziger framework}},\ }\href
  {https://doi.org/10.1103/lds6-34v8} {\bibfield  {journal} {\bibinfo
  {journal} {Phys. Rev. D}\ }\textbf {\bibinfo {volume} {112}},\ \bibinfo
  {pages} {014014} (\bibinfo {year} {2025})},\ \Eprint
  {https://arxiv.org/abs/2505.10602} {arXiv:2505.10602 [hep-th]} \BibitemShut
  {NoStop}%
\bibitem [{\citenamefont {Dudal}\ \emph {et~al.}(2005)\citenamefont {Dudal},
  \citenamefont {Sobreiro}, \citenamefont {Sorella},\ and\ \citenamefont
  {Verschelde}}]{Dudal:2005na}%
  \BibitemOpen
  \bibfield  {author} {\bibinfo {author} {\bibfnamefont {D.}~\bibnamefont
  {Dudal}}, \bibinfo {author} {\bibfnamefont {R.~F.}\ \bibnamefont {Sobreiro}},
  \bibinfo {author} {\bibfnamefont {S.~P.}\ \bibnamefont {Sorella}},\ and\
  \bibinfo {author} {\bibfnamefont {H.}~\bibnamefont {Verschelde}},\ }\bibfield
   {title} {\bibinfo {title} {{The Gribov parameter and the dimension two gluon
  condensate in Euclidean Yang-Mills theories in the Landau gauge}},\ }\href
  {https://doi.org/10.1103/PhysRevD.72.014016} {\bibfield  {journal} {\bibinfo
  {journal} {Phys. Rev. D}\ }\textbf {\bibinfo {volume} {72}},\ \bibinfo
  {pages} {014016} (\bibinfo {year} {2005})},\ \Eprint
  {https://arxiv.org/abs/hep-th/0502183} {arXiv:hep-th/0502183} \BibitemShut
  {NoStop}%
\bibitem [{\citenamefont {Dudal}\ \emph {et~al.}(2019)\citenamefont {Dudal},
  \citenamefont {Felix}, \citenamefont {Palhares}, \citenamefont {Rondeau},\
  and\ \citenamefont {Vercauteren}}]{Dudal:2019ing}%
  \BibitemOpen
  \bibfield  {author} {\bibinfo {author} {\bibfnamefont {D.}~\bibnamefont
  {Dudal}}, \bibinfo {author} {\bibfnamefont {C.~P.}\ \bibnamefont {Felix}},
  \bibinfo {author} {\bibfnamefont {L.~F.}\ \bibnamefont {Palhares}}, \bibinfo
  {author} {\bibfnamefont {F.}~\bibnamefont {Rondeau}},\ and\ \bibinfo {author}
  {\bibfnamefont {D.}~\bibnamefont {Vercauteren}},\ }\bibfield  {title}
  {\bibinfo {title} {{The BRST-invariant vacuum state of the
  Gribov{\textendash}Zwanziger theory}},\ }\href
  {https://doi.org/10.1140/epjc/s10052-019-7235-0} {\bibfield  {journal}
  {\bibinfo  {journal} {Eur. Phys. J. C}\ }\textbf {\bibinfo {volume} {79}},\
  \bibinfo {pages} {731} (\bibinfo {year} {2019})},\ \Eprint
  {https://arxiv.org/abs/1901.11264} {arXiv:1901.11264 [hep-th]} \BibitemShut
  {NoStop}%
\bibitem [{\citenamefont {Capri}\ \emph
  {et~al.}(2013{\natexlab{a}})\citenamefont {Capri}, \citenamefont {Dudal},
  \citenamefont {G\'omez}, \citenamefont {Guimaraes}, \citenamefont {Justo},
  \citenamefont {Sorella},\ and\ \citenamefont {Vercauteren}}]{Capri:2012jhc}%
  \BibitemOpen
  \bibfield  {author} {\bibinfo {author} {\bibfnamefont {M.~A.~L.}\
  \bibnamefont {Capri}}, \bibinfo {author} {\bibfnamefont {D.}~\bibnamefont
  {Dudal}}, \bibinfo {author} {\bibfnamefont {A.~J.}\ \bibnamefont {G\'omez}},
  \bibinfo {author} {\bibfnamefont {M.~S.}\ \bibnamefont {Guimaraes}}, \bibinfo
  {author} {\bibfnamefont {I.~F.}\ \bibnamefont {Justo}}, \bibinfo {author}
  {\bibfnamefont {S.~P.}\ \bibnamefont {Sorella}},\ and\ \bibinfo {author}
  {\bibfnamefont {D.}~\bibnamefont {Vercauteren}},\ }\bibfield  {title}
  {\bibinfo {title} {{Semiclassical analysis of the phases of $4d$ $SU(2)$
  Higgs gauge systems with cutoff at the Gribov horizon}},\ }\href
  {https://doi.org/10.1103/PhysRevD.88.085022} {\bibfield  {journal} {\bibinfo
  {journal} {Phys. Rev. D}\ }\textbf {\bibinfo {volume} {88}},\ \bibinfo
  {pages} {085022} (\bibinfo {year} {2013}{\natexlab{a}})}\BibitemShut
  {NoStop}%
\bibitem [{\citenamefont {Capri}\ \emph {et~al.}(2014)\citenamefont {Capri},
  \citenamefont {Dudal}, \citenamefont {Guimaraes}, \citenamefont {Justo},
  \citenamefont {Sorella},\ and\ \citenamefont {Vercauteren}}]{Capri:2014jhb}%
  \BibitemOpen
  \bibfield  {author} {\bibinfo {author} {\bibfnamefont {M.~A.~L.}\
  \bibnamefont {Capri}}, \bibinfo {author} {\bibfnamefont {D.}~\bibnamefont
  {Dudal}}, \bibinfo {author} {\bibfnamefont {M.~S.}\ \bibnamefont
  {Guimaraes}}, \bibinfo {author} {\bibfnamefont {I.~F.}\ \bibnamefont
  {Justo}}, \bibinfo {author} {\bibfnamefont {S.~P.}\ \bibnamefont {Sorella}},\
  and\ \bibinfo {author} {\bibfnamefont {D.}~\bibnamefont {Vercauteren}},\
  }\bibfield  {title} {\bibinfo {title} {{The (IR-)relevance of the Gribov
  ambiguity in SU(2) x U(1) gauge theories with fundamental Higgs matter}},\
  }\href {https://doi.org/10.1016/j.aop.2014.01.014} {\bibfield  {journal}
  {\bibinfo  {journal} {Annals Phys.}\ }\textbf {\bibinfo {volume} {343}},\
  \bibinfo {pages} {72} (\bibinfo {year} {2014})},\ \Eprint
  {https://arxiv.org/abs/1309.1402} {arXiv:1309.1402 [hep-th]} \BibitemShut
  {NoStop}%
\bibitem [{\citenamefont {Capri}\ \emph
  {et~al.}(2013{\natexlab{b}})\citenamefont {Capri}, \citenamefont {Dudal},
  \citenamefont {Gomez}, \citenamefont {Guimaraes}, \citenamefont {Justo},\
  and\ \citenamefont {Sorella}}]{Capri:2012cr}%
  \BibitemOpen
  \bibfield  {author} {\bibinfo {author} {\bibfnamefont {M.~A.~L.}\
  \bibnamefont {Capri}}, \bibinfo {author} {\bibfnamefont {D.}~\bibnamefont
  {Dudal}}, \bibinfo {author} {\bibfnamefont {A.~J.}\ \bibnamefont {Gomez}},
  \bibinfo {author} {\bibfnamefont {M.~S.}\ \bibnamefont {Guimaraes}}, \bibinfo
  {author} {\bibfnamefont {I.~F.}\ \bibnamefont {Justo}},\ and\ \bibinfo
  {author} {\bibfnamefont {S.~P.}\ \bibnamefont {Sorella}},\ }\bibfield
  {title} {\bibinfo {title} {{A study of the Higgs and confining phases in
  Euclidean SU(2) Yang-Mills theories in 3d by taking into account the Gribov
  horizon}},\ }\href {https://doi.org/10.1140/epjc/s10052-013-2346-5}
  {\bibfield  {journal} {\bibinfo  {journal} {Eur. Phys. J. C}\ }\textbf
  {\bibinfo {volume} {73}},\ \bibinfo {pages} {2346} (\bibinfo {year}
  {2013}{\natexlab{b}})},\ \Eprint {https://arxiv.org/abs/1210.4734}
  {arXiv:1210.4734 [hep-th]} \BibitemShut {NoStop}%
\bibitem [{\citenamefont {Capri}\ \emph
  {et~al.}(2013{\natexlab{c}})\citenamefont {Capri}, \citenamefont {Dudal},
  \citenamefont {Guimaraes}, \citenamefont {Justo}, \citenamefont {Sorella},\
  and\ \citenamefont {Vercauteren}}]{Capri:2013gha}%
  \BibitemOpen
  \bibfield  {author} {\bibinfo {author} {\bibfnamefont {M.~A.~L.}\
  \bibnamefont {Capri}}, \bibinfo {author} {\bibfnamefont {D.}~\bibnamefont
  {Dudal}}, \bibinfo {author} {\bibfnamefont {M.~S.}\ \bibnamefont
  {Guimaraes}}, \bibinfo {author} {\bibfnamefont {I.~F.}\ \bibnamefont
  {Justo}}, \bibinfo {author} {\bibfnamefont {S.~P.}\ \bibnamefont {Sorella}},\
  and\ \bibinfo {author} {\bibfnamefont {D.}~\bibnamefont {Vercauteren}},\
  }\bibfield  {title} {\bibinfo {title} {{$SU(2) \times U(1)$ Yang-Mills
  theories in $3d$ with Higgs field and Gribov ambiguity}},\ }\href
  {https://doi.org/10.1140/epjc/s10052-013-2567-7} {\bibfield  {journal}
  {\bibinfo  {journal} {Eur. Phys. J. C}\ }\textbf {\bibinfo {volume} {73}},\
  \bibinfo {pages} {2567} (\bibinfo {year} {2013}{\natexlab{c}})},\ \Eprint
  {https://arxiv.org/abs/1305.4155} {arXiv:1305.4155 [hep-th]} \BibitemShut
  {NoStop}%
\bibitem [{\citenamefont {Fradkin}\ and\ \citenamefont
  {Shenker}(1979)}]{Fradkin:1979fef}%
  \BibitemOpen
  \bibfield  {author} {\bibinfo {author} {\bibfnamefont {E.}~\bibnamefont
  {Fradkin}}\ and\ \bibinfo {author} {\bibfnamefont {S.~H.}\ \bibnamefont
  {Shenker}},\ }\bibfield  {title} {\bibinfo {title} {{Phase diagrams of
  lattice gauge theories with Higgs fields}},\ }\href
  {https://doi.org/10.1103/PhysRevD.19.3682} {\bibfield  {journal} {\bibinfo
  {journal} {Phys. Rev. D}\ }\textbf {\bibinfo {volume} {19}},\ \bibinfo
  {pages} {3682} (\bibinfo {year} {1979})}\BibitemShut {NoStop}%
\bibitem [{\citenamefont {Canfora}\ \emph {et~al.}(2014)\citenamefont
  {Canfora}, \citenamefont {Gomez}, \citenamefont {Sorella},\ and\
  \citenamefont {Vercauteren}}]{Canfora:2014qwe}%
  \BibitemOpen
  \bibfield  {author} {\bibinfo {author} {\bibfnamefont {F.}~\bibnamefont
  {Canfora}}, \bibinfo {author} {\bibfnamefont {A.}~\bibnamefont {Gomez}},
  \bibinfo {author} {\bibfnamefont {S.~P.}\ \bibnamefont {Sorella}},\ and\
  \bibinfo {author} {\bibfnamefont {D.}~\bibnamefont {Vercauteren}},\
  }\bibfield  {title} {\bibinfo {title} {{Study of
  Yang–Mills–Chern–Simons theory in presence of the Gribov horizon}},\
  }\href {https://doi.org/https://doi.org/10.1016/j.aop.2014.02.017} {\bibfield
   {journal} {\bibinfo  {journal} {Annals of Physics}\ }\textbf {\bibinfo
  {volume} {345}},\ \bibinfo {pages} {166} (\bibinfo {year}
  {2014})}\BibitemShut {NoStop}%
\bibitem [{\citenamefont {Gomez}\ \emph {et~al.}(2016)\citenamefont {Gomez},
  \citenamefont {Gonzalez},\ and\ \citenamefont {Sorella}}]{Gomez:2016egs}%
  \BibitemOpen
  \bibfield  {author} {\bibinfo {author} {\bibfnamefont {A.~J.}\ \bibnamefont
  {Gomez}}, \bibinfo {author} {\bibfnamefont {S.}~\bibnamefont {Gonzalez}},\
  and\ \bibinfo {author} {\bibfnamefont {S.~P.}\ \bibnamefont {Sorella}},\
  }\bibfield  {title} {\bibinfo {title} {Yang–mills–chern–simons system
  in the presence of a gribov horizon with fundamental higgs matter},\ }\href
  {https://doi.org/10.1088/1751-8113/49/6/065401} {\bibfield  {journal}
  {\bibinfo  {journal} {Journal of Physics A: Mathematical and Theoretical}\
  }\textbf {\bibinfo {volume} {49}},\ \bibinfo {pages} {065401} (\bibinfo
  {year} {2016})}\BibitemShut {NoStop}%
\bibitem [{\citenamefont {Ferreira}\ \emph {et~al.}(2020)\citenamefont
  {Ferreira}, \citenamefont {Pereira},\ and\ \citenamefont
  {Sobreiro}}]{Ferreira:2020hcj}%
  \BibitemOpen
  \bibfield  {author} {\bibinfo {author} {\bibfnamefont {L.~C.}\ \bibnamefont
  {Ferreira}}, \bibinfo {author} {\bibfnamefont {A.~D.}\ \bibnamefont
  {Pereira}},\ and\ \bibinfo {author} {\bibfnamefont {R.~F.}\ \bibnamefont
  {Sobreiro}},\ }\bibfield  {title} {\bibinfo {title} {{Hints of confinement
  and deconfinement in Yang-Mills-Chern-Simons theories in the maximal Abelian
  gauge}},\ }\href {https://doi.org/10.1103/PhysRevD.101.105022} {\bibfield
  {journal} {\bibinfo  {journal} {Phys. Rev. D}\ }\textbf {\bibinfo {volume}
  {101}},\ \bibinfo {pages} {105022} (\bibinfo {year} {2020})}\BibitemShut
  {NoStop}%
\bibitem [{\citenamefont {Felix}\ and\ \citenamefont
  {Kao}(2022)}]{Felix:2021eoq}%
  \BibitemOpen
  \bibfield  {author} {\bibinfo {author} {\bibfnamefont {C.~P.}\ \bibnamefont
  {Felix}}\ and\ \bibinfo {author} {\bibfnamefont {C.~W.}\ \bibnamefont
  {Kao}},\ }\bibfield  {title} {\bibinfo {title} {{Implications of the
  topological Chern-Simons mass in the gap equation}},\ }\href
  {https://doi.org/10.1103/PhysRevD.105.094016} {\bibfield  {journal} {\bibinfo
   {journal} {Phys. Rev. D}\ }\textbf {\bibinfo {volume} {105}},\ \bibinfo
  {pages} {094016} (\bibinfo {year} {2022})},\ \Eprint
  {https://arxiv.org/abs/2105.09133} {arXiv:2105.09133 [hep-th]} \BibitemShut
  {NoStop}%
\bibitem [{\citenamefont {Ferreira}\ \emph {et~al.}(2021)\citenamefont
  {Ferreira}, \citenamefont {Granado}, \citenamefont {Justo},\ and\
  \citenamefont {Pereira}}]{Ferreira:2021ksd}%
  \BibitemOpen
  \bibfield  {author} {\bibinfo {author} {\bibfnamefont {L.~C.}\ \bibnamefont
  {Ferreira}}, \bibinfo {author} {\bibfnamefont {D.~R.}\ \bibnamefont
  {Granado}}, \bibinfo {author} {\bibfnamefont {I.~F.}\ \bibnamefont {Justo}},\
  and\ \bibinfo {author} {\bibfnamefont {A.~D.}\ \bibnamefont {Pereira}},\
  }\bibfield  {title} {\bibinfo {title} {{Infrared propagators of
  Yang-Mills-Chern-Simons theories in linear covariant gauges}},\ }\href
  {https://doi.org/10.1103/PhysRevD.104.045007} {\bibfield  {journal} {\bibinfo
   {journal} {Phys. Rev. D}\ }\textbf {\bibinfo {volume} {104}},\ \bibinfo
  {pages} {045007} (\bibinfo {year} {2021})}\BibitemShut {NoStop}%
\bibitem [{\citenamefont {Deser}\ \emph
  {et~al.}(1982{\natexlab{a}})\citenamefont {Deser}, \citenamefont {Jackiw},\
  and\ \citenamefont {Templeton}}]{Deser:1982fsr}%
  \BibitemOpen
  \bibfield  {author} {\bibinfo {author} {\bibfnamefont {S.}~\bibnamefont
  {Deser}}, \bibinfo {author} {\bibfnamefont {R.}~\bibnamefont {Jackiw}},\ and\
  \bibinfo {author} {\bibfnamefont {S.}~\bibnamefont {Templeton}},\ }\bibfield
  {title} {\bibinfo {title} {{Topologically massive gauge theories}},\ }\href
  {https://doi.org/https://doi.org/10.1016/0003-4916(82)90164-6} {\bibfield
  {journal} {\bibinfo  {journal} {Annals of Physics}\ }\textbf {\bibinfo
  {volume} {140}},\ \bibinfo {pages} {372} (\bibinfo {year}
  {1982}{\natexlab{a}})}\BibitemShut {NoStop}%
\bibitem [{\citenamefont {Deser}\ \emph
  {et~al.}(1982{\natexlab{b}})\citenamefont {Deser}, \citenamefont {Jackiw},\
  and\ \citenamefont {Templeton}}]{Deser:1982kay}%
  \BibitemOpen
  \bibfield  {author} {\bibinfo {author} {\bibfnamefont {S.}~\bibnamefont
  {Deser}}, \bibinfo {author} {\bibfnamefont {R.}~\bibnamefont {Jackiw}},\ and\
  \bibinfo {author} {\bibfnamefont {S.}~\bibnamefont {Templeton}},\ }\bibfield
  {title} {\bibinfo {title} {{Three-Dimensional Massive Gauge Theories}},\
  }\href {https://doi.org/10.1103/PhysRevLett.48.975} {\bibfield  {journal}
  {\bibinfo  {journal} {Phys. Rev. Lett.}\ }\textbf {\bibinfo {volume} {48}},\
  \bibinfo {pages} {975} (\bibinfo {year} {1982}{\natexlab{b}})}\BibitemShut
  {NoStop}%
\bibitem [{\citenamefont {Delduc}\ \emph {et~al.}(1990)\citenamefont {Delduc},
  \citenamefont {Lucchesi}, \citenamefont {Piguet},\ and\ \citenamefont
  {Sorella}}]{Delduc:1990je}%
  \BibitemOpen
  \bibfield  {author} {\bibinfo {author} {\bibfnamefont {F.}~\bibnamefont
  {Delduc}}, \bibinfo {author} {\bibfnamefont {C.}~\bibnamefont {Lucchesi}},
  \bibinfo {author} {\bibfnamefont {O.}~\bibnamefont {Piguet}},\ and\ \bibinfo
  {author} {\bibfnamefont {S.~P.}\ \bibnamefont {Sorella}},\ }\bibfield
  {title} {\bibinfo {title} {{Exact Scale Invariance of the {Chern-Simons}
  Theory in the Landau Gauge}},\ }\href
  {https://doi.org/10.1016/0550-3213(90)90283-J} {\bibfield  {journal}
  {\bibinfo  {journal} {Nucl. Phys. B}\ }\textbf {\bibinfo {volume} {346}},\
  \bibinfo {pages} {313} (\bibinfo {year} {1990})}\BibitemShut {NoStop}%
\bibitem [{\citenamefont {Giavarini}\ \emph {et~al.}(1992)\citenamefont
  {Giavarini}, \citenamefont {Martin},\ and\ \citenamefont
  {Ruiz~Ruiz}}]{Giavarini:1992xz}%
  \BibitemOpen
  \bibfield  {author} {\bibinfo {author} {\bibfnamefont {G.}~\bibnamefont
  {Giavarini}}, \bibinfo {author} {\bibfnamefont {C.~P.}\ \bibnamefont
  {Martin}},\ and\ \bibinfo {author} {\bibfnamefont {F.}~\bibnamefont
  {Ruiz~Ruiz}},\ }\bibfield  {title} {\bibinfo {title} {{Chern-Simons theory as
  the large mass limit of topologically massive Yang-Mills theory}},\ }\href
  {https://doi.org/10.1016/0550-3213(92)90647-T} {\bibfield  {journal}
  {\bibinfo  {journal} {Nucl. Phys. B}\ }\textbf {\bibinfo {volume} {381}},\
  \bibinfo {pages} {222} (\bibinfo {year} {1992})},\ \Eprint
  {https://arxiv.org/abs/hep-th/9206007} {arXiv:hep-th/9206007} \BibitemShut
  {NoStop}%
\bibitem [{\citenamefont {Del~Cima}\ \emph {et~al.}(1998)\citenamefont
  {Del~Cima}, \citenamefont {Franco}, \citenamefont {Helayel-Neto},\ and\
  \citenamefont {Piguet}}]{DelCima:1997pb}%
  \BibitemOpen
  \bibfield  {author} {\bibinfo {author} {\bibfnamefont {O.~M.}\ \bibnamefont
  {Del~Cima}}, \bibinfo {author} {\bibfnamefont {D.~H.~T.}\ \bibnamefont
  {Franco}}, \bibinfo {author} {\bibfnamefont {J.~A.}\ \bibnamefont
  {Helayel-Neto}},\ and\ \bibinfo {author} {\bibfnamefont {O.}~\bibnamefont
  {Piguet}},\ }\bibfield  {title} {\bibinfo {title} {{On the nonrenormalization
  properties of gauge theories with Chern-Simons terms}},\ }\href
  {https://doi.org/10.1088/1126-6708/1998/02/002} {\bibfield  {journal}
  {\bibinfo  {journal} {JHEP}\ }\textbf {\bibinfo {volume} {02}},\ \bibinfo
  {pages} {002}},\ \Eprint {https://arxiv.org/abs/hep-th/9711191}
  {arXiv:hep-th/9711191} \BibitemShut {NoStop}%
\bibitem [{\citenamefont {Del~Cima}\ \emph {et~al.}(1999)\citenamefont
  {Del~Cima}, \citenamefont {Franco}, \citenamefont {Helayel-Neto},\ and\
  \citenamefont {Piguet}}]{DelCima:1998ur}%
  \BibitemOpen
  \bibfield  {author} {\bibinfo {author} {\bibfnamefont {O.~M.}\ \bibnamefont
  {Del~Cima}}, \bibinfo {author} {\bibfnamefont {D.~H.~T.}\ \bibnamefont
  {Franco}}, \bibinfo {author} {\bibfnamefont {J.~A.}\ \bibnamefont
  {Helayel-Neto}},\ and\ \bibinfo {author} {\bibfnamefont {O.}~\bibnamefont
  {Piguet}},\ }\bibfield  {title} {\bibinfo {title} {{An Algebraic proof on the
  finiteness of Yang-Mills-Chern-Simons theory in D = 3}},\ }\href
  {https://doi.org/10.1023/A:1007595121742} {\bibfield  {journal} {\bibinfo
  {journal} {Lett. Math. Phys.}\ }\textbf {\bibinfo {volume} {47}},\ \bibinfo
  {pages} {265} (\bibinfo {year} {1999})},\ \Eprint
  {https://arxiv.org/abs/math-ph/9904030} {arXiv:math-ph/9904030} \BibitemShut
  {NoStop}%
\bibitem [{\citenamefont {Azevedo}\ \emph {et~al.}(2024)\citenamefont
  {Azevedo}, \citenamefont {Del~Cima}, \citenamefont {Dias},\ and\
  \citenamefont {Pereira}}]{Azevedo:2024cov}%
  \BibitemOpen
  \bibfield  {author} {\bibinfo {author} {\bibfnamefont {D.~O.~R.}\
  \bibnamefont {Azevedo}}, \bibinfo {author} {\bibfnamefont {O.~M.}\
  \bibnamefont {Del~Cima}}, \bibinfo {author} {\bibfnamefont {T.~S.}\
  \bibnamefont {Dias}},\ and\ \bibinfo {author} {\bibfnamefont {E.~D.}\
  \bibnamefont {Pereira}},\ }\bibfield  {title} {\bibinfo {title}
  {{Interpolating gauge-fixing for
  Yang{\textendash}Mills{\textendash}Chern{\textendash}Simons theory in
  $D=3$}},\ }\href {https://doi.org/10.1140/epjc/s10052-024-13279-3} {\bibfield
   {journal} {\bibinfo  {journal} {Eur. Phys. J. C}\ }\textbf {\bibinfo
  {volume} {84}},\ \bibinfo {pages} {917} (\bibinfo {year} {2024})},\ \Eprint
  {https://arxiv.org/abs/2406.09515} {arXiv:2406.09515 [hep-th]} \BibitemShut
  {NoStop}%
\bibitem [{\citenamefont {Azevedo}\ and\ \citenamefont
  {Pereira}(2025)}]{Azevedo:2025yal}%
  \BibitemOpen
  \bibfield  {author} {\bibinfo {author} {\bibfnamefont {D.~O.~R.}\
  \bibnamefont {Azevedo}}\ and\ \bibinfo {author} {\bibfnamefont {A.~D.}\
  \bibnamefont {Pereira}},\ }\bibfield  {title} {\bibinfo {title} {{Finiteness
  of the Yang-Mills-Chern-Simons action in linear covariant gauges by taking
  into account gauge copies}},\ }\href
  {https://doi.org/10.1103/PhysRevD.111.085028} {\bibfield  {journal} {\bibinfo
   {journal} {Phys. Rev. D}\ }\textbf {\bibinfo {volume} {111}},\ \bibinfo
  {pages} {085028} (\bibinfo {year} {2025})},\ \Eprint
  {https://arxiv.org/abs/2502.03284} {arXiv:2502.03284 [hep-th]} \BibitemShut
  {NoStop}%
\bibitem [{\citenamefont {Capri}\ \emph
  {et~al.}(2013{\natexlab{d}})\citenamefont {Capri}, \citenamefont {Dudal},
  \citenamefont {Guimaraes}, \citenamefont {Palhares},\ and\ \citenamefont
  {Sorella}}]{Capri:2012wx}%
  \BibitemOpen
  \bibfield  {author} {\bibinfo {author} {\bibfnamefont {M.~A.~L.}\
  \bibnamefont {Capri}}, \bibinfo {author} {\bibfnamefont {D.}~\bibnamefont
  {Dudal}}, \bibinfo {author} {\bibfnamefont {M.~S.}\ \bibnamefont
  {Guimaraes}}, \bibinfo {author} {\bibfnamefont {L.~F.}\ \bibnamefont
  {Palhares}},\ and\ \bibinfo {author} {\bibfnamefont {S.~P.}\ \bibnamefont
  {Sorella}},\ }\bibfield  {title} {\bibinfo {title} {{An all-order proof of
  the equivalence between Gribov's no-pole and Zwanziger's horizon
  conditions}},\ }\href {https://doi.org/10.1016/j.physletb.2013.01.039}
  {\bibfield  {journal} {\bibinfo  {journal} {Phys. Lett.}\ }\textbf {\bibinfo
  {volume} {B719}},\ \bibinfo {pages} {448} (\bibinfo {year}
  {2013}{\natexlab{d}})},\ \Eprint {https://arxiv.org/abs/1212.2419}
  {arXiv:1212.2419 [hep-th]} \BibitemShut {NoStop}%
\bibitem [{\citenamefont {Capri}\ \emph {et~al.}(2015)\citenamefont {Capri},
  \citenamefont {Dudal}, \citenamefont {Fiorentini}, \citenamefont {Guimaraes},
  \citenamefont {Justo}, \citenamefont {Pereira}, \citenamefont {Mintz},
  \citenamefont {Palhares}, \citenamefont {Sobreiro},\ and\ \citenamefont
  {Sorella}}]{Capri:2015ixa}%
  \BibitemOpen
  \bibfield  {author} {\bibinfo {author} {\bibfnamefont {M.~A.~L.}\
  \bibnamefont {Capri}}, \bibinfo {author} {\bibfnamefont {D.}~\bibnamefont
  {Dudal}}, \bibinfo {author} {\bibfnamefont {D.}~\bibnamefont {Fiorentini}},
  \bibinfo {author} {\bibfnamefont {M.~S.}\ \bibnamefont {Guimaraes}}, \bibinfo
  {author} {\bibfnamefont {I.~F.}\ \bibnamefont {Justo}}, \bibinfo {author}
  {\bibfnamefont {A.~D.}\ \bibnamefont {Pereira}}, \bibinfo {author}
  {\bibfnamefont {B.~W.}\ \bibnamefont {Mintz}}, \bibinfo {author}
  {\bibfnamefont {L.~F.}\ \bibnamefont {Palhares}}, \bibinfo {author}
  {\bibfnamefont {R.~F.}\ \bibnamefont {Sobreiro}},\ and\ \bibinfo {author}
  {\bibfnamefont {S.~P.}\ \bibnamefont {Sorella}},\ }\bibfield  {title}
  {\bibinfo {title} {{Exact nilpotent nonperturbative BRST symmetry for the
  Gribov-Zwanziger action in the linear covariant gauge}},\ }\href
  {https://doi.org/10.1103/PhysRevD.92.045039} {\bibfield  {journal} {\bibinfo
  {journal} {Phys. Rev. D}\ }\textbf {\bibinfo {volume} {92}},\ \bibinfo
  {pages} {045039} (\bibinfo {year} {2015})},\ \Eprint
  {https://arxiv.org/abs/1506.06995} {arXiv:1506.06995 [hep-th]} \BibitemShut
  {NoStop}%
\bibitem [{\citenamefont {Capri}\ \emph
  {et~al.}(2016{\natexlab{a}})\citenamefont {Capri}, \citenamefont
  {Fiorentini}, \citenamefont {Guimaraes}, \citenamefont {Mintz}, \citenamefont
  {Palhares}, \citenamefont {Sorella}, \citenamefont {Dudal}, \citenamefont
  {Justo}, \citenamefont {Pereira},\ and\ \citenamefont
  {Sobreiro}}]{Capri:2015nzw}%
  \BibitemOpen
  \bibfield  {author} {\bibinfo {author} {\bibfnamefont {M.~A.~L.}\
  \bibnamefont {Capri}}, \bibinfo {author} {\bibfnamefont {D.}~\bibnamefont
  {Fiorentini}}, \bibinfo {author} {\bibfnamefont {M.~S.}\ \bibnamefont
  {Guimaraes}}, \bibinfo {author} {\bibfnamefont {B.~W.}\ \bibnamefont
  {Mintz}}, \bibinfo {author} {\bibfnamefont {L.~F.}\ \bibnamefont {Palhares}},
  \bibinfo {author} {\bibfnamefont {S.~P.}\ \bibnamefont {Sorella}}, \bibinfo
  {author} {\bibfnamefont {D.}~\bibnamefont {Dudal}}, \bibinfo {author}
  {\bibfnamefont {I.~F.}\ \bibnamefont {Justo}}, \bibinfo {author}
  {\bibfnamefont {A.~D.}\ \bibnamefont {Pereira}},\ and\ \bibinfo {author}
  {\bibfnamefont {R.~F.}\ \bibnamefont {Sobreiro}},\ }\bibfield  {title}
  {\bibinfo {title} {{More on the nonperturbative Gribov-Zwanziger quantization
  of linear covariant gauges}},\ }\href
  {https://doi.org/10.1103/PhysRevD.93.065019} {\bibfield  {journal} {\bibinfo
  {journal} {Phys. Rev. D}\ }\textbf {\bibinfo {volume} {93}},\ \bibinfo
  {pages} {065019} (\bibinfo {year} {2016}{\natexlab{a}})},\ \Eprint
  {https://arxiv.org/abs/1512.05833} {arXiv:1512.05833 [hep-th]} \BibitemShut
  {NoStop}%
\bibitem [{\citenamefont {Capri}\ \emph
  {et~al.}(2016{\natexlab{b}})\citenamefont {Capri}, \citenamefont {Dudal},
  \citenamefont {Fiorentini}, \citenamefont {Guimaraes}, \citenamefont {Justo},
  \citenamefont {Pereira}, \citenamefont {Mintz}, \citenamefont {Palhares},
  \citenamefont {Sobreiro},\ and\ \citenamefont {Sorella}}]{Capri:2016aqq}%
  \BibitemOpen
  \bibfield  {author} {\bibinfo {author} {\bibfnamefont {M.~A.~L.}\
  \bibnamefont {Capri}}, \bibinfo {author} {\bibfnamefont {D.}~\bibnamefont
  {Dudal}}, \bibinfo {author} {\bibfnamefont {D.}~\bibnamefont {Fiorentini}},
  \bibinfo {author} {\bibfnamefont {M.~S.}\ \bibnamefont {Guimaraes}}, \bibinfo
  {author} {\bibfnamefont {I.~F.}\ \bibnamefont {Justo}}, \bibinfo {author}
  {\bibfnamefont {A.~D.}\ \bibnamefont {Pereira}}, \bibinfo {author}
  {\bibfnamefont {B.~W.}\ \bibnamefont {Mintz}}, \bibinfo {author}
  {\bibfnamefont {L.~F.}\ \bibnamefont {Palhares}}, \bibinfo {author}
  {\bibfnamefont {R.~F.}\ \bibnamefont {Sobreiro}},\ and\ \bibinfo {author}
  {\bibfnamefont {S.~P.}\ \bibnamefont {Sorella}},\ }\bibfield  {title}
  {\bibinfo {title} {{Local and BRST-invariant Yang-Mills theory within the
  Gribov horizon}},\ }\href {https://doi.org/10.1103/PhysRevD.94.025035}
  {\bibfield  {journal} {\bibinfo  {journal} {Phys. Rev. D}\ }\textbf {\bibinfo
  {volume} {94}},\ \bibinfo {pages} {025035} (\bibinfo {year}
  {2016}{\natexlab{b}})},\ \Eprint {https://arxiv.org/abs/1605.02610}
  {arXiv:1605.02610 [hep-th]} \BibitemShut {NoStop}%
\bibitem [{\citenamefont {Capri}\ \emph
  {et~al.}(2017{\natexlab{a}})\citenamefont {Capri}, \citenamefont {Dudal},
  \citenamefont {Pereira}, \citenamefont {Fiorentini}, \citenamefont
  {Guimaraes}, \citenamefont {Mintz}, \citenamefont {Palhares},\ and\
  \citenamefont {Sorella}}]{Capri:2016gut}%
  \BibitemOpen
  \bibfield  {author} {\bibinfo {author} {\bibfnamefont {M.~A.~L.}\
  \bibnamefont {Capri}}, \bibinfo {author} {\bibfnamefont {D.}~\bibnamefont
  {Dudal}}, \bibinfo {author} {\bibfnamefont {A.~D.}\ \bibnamefont {Pereira}},
  \bibinfo {author} {\bibfnamefont {D.}~\bibnamefont {Fiorentini}}, \bibinfo
  {author} {\bibfnamefont {M.~S.}\ \bibnamefont {Guimaraes}}, \bibinfo {author}
  {\bibfnamefont {B.~W.}\ \bibnamefont {Mintz}}, \bibinfo {author}
  {\bibfnamefont {L.~F.}\ \bibnamefont {Palhares}},\ and\ \bibinfo {author}
  {\bibfnamefont {S.~P.}\ \bibnamefont {Sorella}},\ }\bibfield  {title}
  {\bibinfo {title} {{Nonperturbative aspects of Euclidean Yang-Mills theories
  in linear covariant gauges: Nielsen identities and a BRST-invariant two-point
  correlation function}},\ }\href {https://doi.org/10.1103/PhysRevD.95.045011}
  {\bibfield  {journal} {\bibinfo  {journal} {Phys. Rev. D}\ }\textbf {\bibinfo
  {volume} {95}},\ \bibinfo {pages} {045011} (\bibinfo {year}
  {2017}{\natexlab{a}})},\ \Eprint {https://arxiv.org/abs/1611.10077}
  {arXiv:1611.10077 [hep-th]} \BibitemShut {NoStop}%
\bibitem [{\citenamefont {Capri}\ \emph
  {et~al.}(2017{\natexlab{b}})\citenamefont {Capri}, \citenamefont
  {Fiorentini}, \citenamefont {Pereira},\ and\ \citenamefont
  {Sorella}}]{Capri:2017bfd}%
  \BibitemOpen
  \bibfield  {author} {\bibinfo {author} {\bibfnamefont {M.~A.~L.}\
  \bibnamefont {Capri}}, \bibinfo {author} {\bibfnamefont {D.}~\bibnamefont
  {Fiorentini}}, \bibinfo {author} {\bibfnamefont {A.~D.}\ \bibnamefont
  {Pereira}},\ and\ \bibinfo {author} {\bibfnamefont {S.~P.}\ \bibnamefont
  {Sorella}},\ }\bibfield  {title} {\bibinfo {title} {{Renormalizability of the
  refined Gribov-Zwanziger action in linear covariant gauges}},\ }\href
  {https://doi.org/10.1103/PhysRevD.96.054022} {\bibfield  {journal} {\bibinfo
  {journal} {Phys. Rev. D}\ }\textbf {\bibinfo {volume} {96}},\ \bibinfo
  {pages} {054022} (\bibinfo {year} {2017}{\natexlab{b}})},\ \Eprint
  {https://arxiv.org/abs/1708.01543} {arXiv:1708.01543 [hep-th]} \BibitemShut
  {NoStop}%
\bibitem [{\citenamefont {Capri}\ \emph {et~al.}(2018)\citenamefont {Capri},
  \citenamefont {Dudal}, \citenamefont {Guimaraes}, \citenamefont {Pereira},
  \citenamefont {Mintz}, \citenamefont {Palhares},\ and\ \citenamefont
  {Sorella}}]{Capri:2018ijg}%
  \BibitemOpen
  \bibfield  {author} {\bibinfo {author} {\bibfnamefont {M.~A.~L.}\
  \bibnamefont {Capri}}, \bibinfo {author} {\bibfnamefont {D.}~\bibnamefont
  {Dudal}}, \bibinfo {author} {\bibfnamefont {M.~S.}\ \bibnamefont
  {Guimaraes}}, \bibinfo {author} {\bibfnamefont {A.~D.}\ \bibnamefont
  {Pereira}}, \bibinfo {author} {\bibfnamefont {B.~W.}\ \bibnamefont {Mintz}},
  \bibinfo {author} {\bibfnamefont {L.~F.}\ \bibnamefont {Palhares}},\ and\
  \bibinfo {author} {\bibfnamefont {S.~P.}\ \bibnamefont {Sorella}},\
  }\bibfield  {title} {\bibinfo {title} {{The universal character of
  Zwanziger's horizon function in Euclidean Yang\textendash{}Mills theories}},\
  }\href {https://doi.org/10.1016/j.physletb.2018.03.058} {\bibfield  {journal}
  {\bibinfo  {journal} {Phys. Lett. B}\ }\textbf {\bibinfo {volume} {781}},\
  \bibinfo {pages} {48} (\bibinfo {year} {2018})},\ \Eprint
  {https://arxiv.org/abs/1802.04582} {arXiv:1802.04582 [hep-th]} \BibitemShut
  {NoStop}%
\bibitem [{\citenamefont {Capri}\ \emph
  {et~al.}(2017{\natexlab{c}})\citenamefont {Capri}, \citenamefont
  {Fiorentini}, \citenamefont {Pereira},\ and\ \citenamefont
  {Sorella}}]{Capri:2017abz}%
  \BibitemOpen
  \bibfield  {author} {\bibinfo {author} {\bibfnamefont {M.~A.~L.}\
  \bibnamefont {Capri}}, \bibinfo {author} {\bibfnamefont {D.}~\bibnamefont
  {Fiorentini}}, \bibinfo {author} {\bibfnamefont {A.~D.}\ \bibnamefont
  {Pereira}},\ and\ \bibinfo {author} {\bibfnamefont {S.~P.}\ \bibnamefont
  {Sorella}},\ }\bibfield  {title} {\bibinfo {title} {{A non-perturbative study
  of matter field propagators in Euclidean Yang\textendash{}Mills theory in
  linear covariant, Curci\textendash{}Ferrari and maximal Abelian gauges}},\
  }\href {https://doi.org/10.1140/epjc/s10052-017-5107-z} {\bibfield  {journal}
  {\bibinfo  {journal} {Eur. Phys. J. C}\ }\textbf {\bibinfo {volume} {77}},\
  \bibinfo {pages} {546} (\bibinfo {year} {2017}{\natexlab{c}})},\ \Eprint
  {https://arxiv.org/abs/1703.03264} {arXiv:1703.03264 [hep-th]} \BibitemShut
  {NoStop}%
\bibitem [{\citenamefont {Zhang}\ and\ \citenamefont
  {Chen}(2025)}]{Zhang:2024sgm}%
  \BibitemOpen
  \bibfield  {author} {\bibinfo {author} {\bibfnamefont {P.}~\bibnamefont
  {Zhang}}\ and\ \bibinfo {author} {\bibfnamefont {J.-Y.}\ \bibnamefont
  {Chen}},\ }\bibfield  {title} {\bibinfo {title} {{An explicit categorical
  construction of instanton density in lattice Yang-Mills theory}},\ }\href
  {https://doi.org/10.1007/JHEP06(2025)085} {\bibfield  {journal} {\bibinfo
  {journal} {JHEP}\ }\textbf {\bibinfo {volume} {06}},\ \bibinfo {pages}
  {085}},\ \Eprint {https://arxiv.org/abs/2411.07195} {arXiv:2411.07195
  [hep-lat]} \BibitemShut {NoStop}%
\bibitem [{\citenamefont {Xu}\ and\ \citenamefont {Chen}(2025)}]{Xu:2024hyo}%
  \BibitemOpen
  \bibfield  {author} {\bibinfo {author} {\bibfnamefont {Z.-A.}\ \bibnamefont
  {Xu}}\ and\ \bibinfo {author} {\bibfnamefont {J.-Y.}\ \bibnamefont {Chen}},\
  }\bibfield  {title} {\bibinfo {title} {{Lattice Chern-Simons-Maxwell theory
  and its chirality}},\ }\href {https://doi.org/10.1007/JHEP08(2025)062}
  {\bibfield  {journal} {\bibinfo  {journal} {JHEP}\ }\textbf {\bibinfo
  {volume} {08}},\ \bibinfo {pages} {062}},\ \Eprint
  {https://arxiv.org/abs/2410.11034} {arXiv:2410.11034 [hep-th]} \BibitemShut
  {NoStop}%
\bibitem [{\citenamefont {Peng}\ \emph {et~al.}(2025)\citenamefont {Peng},
  \citenamefont {Diamantini}, \citenamefont {Funcke}, \citenamefont {Hassan},
  \citenamefont {Jansen}, \citenamefont {K{\"u}hn}, \citenamefont {Luo},\ and\
  \citenamefont {Naredi}}]{Peng:2025nfa}%
  \BibitemOpen
  \bibfield  {author} {\bibinfo {author} {\bibfnamefont {C.}~\bibnamefont
  {Peng}}, \bibinfo {author} {\bibfnamefont {M.~C.}\ \bibnamefont
  {Diamantini}}, \bibinfo {author} {\bibfnamefont {L.}~\bibnamefont {Funcke}},
  \bibinfo {author} {\bibfnamefont {S.~M.~A.}\ \bibnamefont {Hassan}}, \bibinfo
  {author} {\bibfnamefont {K.}~\bibnamefont {Jansen}}, \bibinfo {author}
  {\bibfnamefont {S.}~\bibnamefont {K{\"u}hn}}, \bibinfo {author}
  {\bibfnamefont {D.}~\bibnamefont {Luo}},\ and\ \bibinfo {author}
  {\bibfnamefont {P.}~\bibnamefont {Naredi}},\ }\bibfield  {title} {\bibinfo
  {title} {{Hamiltonian Lattice Formulation of Compact Maxwell-Chern-Simons
  Theory}},\ }\href@noop {} {\  (\bibinfo {year} {2025})},\ \Eprint
  {https://arxiv.org/abs/2407.20225} {arXiv:2407.20225 [hep-th]} \BibitemShut
  {NoStop}%
\end{thebibliography}%

\end{document}